\documentclass[pdflatex,sn-aps]{sn-jnl}

\usepackage{graphicx}%
\usepackage{multirow}%
\usepackage{amsmath,amssymb,amsfonts}%
\usepackage{amsthm}%
\usepackage{mathrsfs}%
\usepackage[title]{appendix}%
\usepackage{xcolor}%
\usepackage{textcomp}%
\usepackage{manyfoot}%
\usepackage{booktabs}%
\usepackage{algorithm}%
\usepackage{algorithmicx}%
\usepackage{algpseudocode}%
\usepackage{listings}%
\usepackage{tabularx}

\theoremstyle{thmstyleone}%
\theoremstyle{thmstyletwo}%

\theoremstyle{thmstylethree}%

\begin{document}

\title[Article Title]{Scale-Dependent Collective Adaptation in Self-Amending LLM Societies: A Cross-Family Study of Emergent Governance}


\author*[1]{\fnm{Kazuya} \sur{Horibe}}\email{kazuya.horibe@riken.com}

\author[2]{\fnm{Masaomi} \sur{Hatakeyama}}

\author[3]{\fnm{Gen} \sur{Masumoto}}

\author[4]{\fnm{Takashi} \sur{Hashimoto}}

\author[5,6]{\fnm{Peter} \sur{Romero}}

\affil*[1]{\orgdiv{Center for Brain Science}, \orgname{RIKEN}}

\affil[2]{\orgdiv{Department of Evolutionary Biology and Environmental Studies}, \orgname{University of Zurich}}

\affil[3]{\orgdiv{Information R\&D and Strategy Headquarters}, \orgname{RIKEN}}

\affil[4]{\orgdiv{School of Knowledge Science}, \orgname{Japan Advanced Institute of Science and Technology}}

\affil[5]{\orgdiv{Valencian Research Institute of Artificial Intelligence}, \orgname{Universitat Polit\`ecnica de Val\`encia}}


\affil[6]{\orgdiv{The Psychometrics Centre}, \orgname{University of Cambridge}}


\abstract{We study group decision-making in artificial societies where the rules of play are themselves subject to collective amendment. Using the self-amending game Nomic, we compare multiple scales across two LLM families and find that collective adaptation does not improve monotonically with model size. Instead, both families exhibit a narrow mid-scale regime that supports sustained rule adoption, diverse amendments, and balanced consensus. Smaller models tend to remain rule-inert, whereas larger models often converge on restrictive voting patterns, and heterogeneous mixed-size groups collapse into veto-driven gridlock. These cross-scale contrasts persist under temperature perturbations and under a shift from unanimity to majority voting, although latent-state structure varies by family and scale. Hidden-state divergence alone does not explain collective performance: high representational divergence can coincide with poor behavioural outcomes. Linear probes reveal regime-selective coupling between latent vote-predictive signals and collective behaviour, but decodability is necessary rather than sufficient for adaptive play. Overall, the recurring regularity is non-monotonicity, not the particular scale at which the optimum appears. Self-amending games therefore provide a controlled testbed for studying collective adaptation in artificial societies beyond raw model scale.}

\keywords{collective decision-making, emergent governance, institutional evolution, agent-based simulation, large language models, cooperation}



\maketitle

\section{Introduction}\label{sec:intro}

When large language model (LLM) societies rewrite their own rules of play, institutional change does not improve monotonically with model scale.
Across two architecturally distinct model families, Qwen3.5~\cite{qwen2026qwen35} and Gemma3~\cite{google2025gemma3}, collective adaptation is non-monotonic: each family produces one mid-scale optimum at a family-specific parameter count, while smaller and larger scales fail in different ways.
The cross-family pattern is the non-monotonic shape itself, not the specific scale at which the sweet spot appears.

Collective adaptation, that is, the coupled evolution of strategies and the institutions that govern them, is widespread in social systems from legislatures to open-source communities~\cite{axelrod1984evolution,ostrom1990governing,north1990institutions,henrich2016secret,boyd2008geneculture,galesic2023beyond}, yet remains difficult to formalise because agents rewrite the same rules they follow.
Formal frameworks for analysing games whose rules are themselves objects of decision have been proposed in the meta-game tradition, including a lambda-calculus formulation that treats rule-changes as operators on the game itself~\cite{masumoto2005new}.
LLMs make these endogenous processes simulable because they can read and generate the natural-language content through which institutional rules are debated and rewritten~\cite{park2023generative,grossmann2023ai,bail2024generative,argyle2023out,zeng2026toohuman}.
Most game-theoretic benchmarks for LLMs still hold the institution fixed: classical matrix games~\cite{cobben2026gt,duan2024gtbench,costarelli2024gamebench}, Diplomacy~\cite{meta2022human}, social deduction and imperfect-information play~\cite{xu2023exploring,light2023avalonbench,guo2023suspicion,hu2024survey}, competition and scaffolded cooperation~\cite{zhao2024competeai,mckee2023scaffolding}, multi-agent debate and collaboration~\cite{du2024improving,li2023camel}, repeated-game studies~\cite{akata2025playing,gandhi2023strategic}, and extended role-playing evaluations~\cite{pan2023rewards,lan2024llm} all measure strategic play under static rules (see~\cite{sun2025game} for a survey).
Agents in these settings can deceive, negotiate, and form coalitions, but the rule set stays fixed, so collective adaptation is out of scope by construction.

Nomic~\cite{suber1990paradox} provides a controlled testbed for collective adaptation because each turn couples a proposal to rewrite a rule with a social vote, and a reduced variant (\emph{Minimum Nomic}) has been used to study rule-dynamics with manageable state spaces~\cite{hatakeyama2009minimum}.
Related work has examined LLM-mediated law-making~\cite{hota2025nomiclaw,huang2024collective}, self-modifying prompt populations~\cite{fernando2023promptbreeder}, self-amending agent swarms~\cite{horibe2025evolvability}, and open-ended or embodied adaptation~\cite{hughes2024open,clune2019ai,wang2023voyager,lehman2023evolution,stanley2015greatness}, alongside alignment pipelines that allow models to rewrite their own outputs~\cite{ouyang2022training,rafailov2023direct,bai2022constitutional,christiano2017deep,perez2023discovering}.
Basic rule-following nevertheless remains fragile in multi-step games~\cite{mu2023can}, and dependence of collective adaptation on model capability is still unresolved.

Scaling is the natural independent variable for this question, but collective adaptation has not been characterised along that axis.
Scaling laws predict monotonic improvement on token-level objectives~\cite{kaplan2020scaling,hoffmann2022training}, and emergent abilities have been catalogued at specific parameter thresholds~\cite{brown2020language,wei2022emergent}, though apparent transitions may reflect metric discontinuities~\cite{schaeffer2023emergent} and U-shaped or inverse-scaling regimes are now well documented~\cite{mckenzie2024inverse,wu2025ushaped}.
Rule-making couples technical correctness (writing a well-typed amendment) with social coordination (getting it voted through), so it remains open whether either axis scales monotonically.
Related capabilities, including chain-of-thought prompting~\cite{wei2022chain,kojima2022large}, tree-structured deliberation~\cite{yao2023tree}, self-consistency and self-refinement~\cite{wang2022self,madaan2023self,shinn2023reflexion}, theory of mind~\cite{kosinski2024evaluating,ullman2023large,sap2022neural}, and metacognitive monitoring~\cite{flavell1979metacognition,didolkar2024metacognitive}, develop unevenly across scale.
This study measures these patterns functionally across both model families, without attributing genuine understanding~\cite{searle1980minds}.

To bring this question under controlled conditions, we vary model scale within both Qwen3.5 (0.8B, 2B, 4B, 9B, 27B) and Gemma3 (270M, 1B, 4B, 12B, 27B), holding architecture and training recipe fixed within each family, and pair behavioural measurement with representational analysis through hidden-state probing~\cite{alain2017understanding,belinkov2022probing,nostalgebraist2020logitlens,belrose2023eliciting,meng2022locating,vig2020investigating,lindsey2025emergent,li2023emergent,gurnee2024language}.
This cross-family design separates shared regularities from architecture-specific effects.
The remainder of the paper instantiates the design in the rule-rewriting game Nomic, which makes the institution itself the object of play.

\section{Results}\label{sec:results}

In each round of Nomic, each of five players takes one turn and proposes a single rule change of type ADD (create a rule), AMEND (modify an existing one), REPEAL (delete one), or TRANSMUTE (flip a rule between immutable and mutable), over a starting constitution of 29 rules (16 immutable, 13 mutable).
The remaining players then vote on the proposal; an adopted proposal rewrites the institution and carries forward.
Points are earned principally through a turn-start dice roll (Rule~202), and a game terminates when a player reaches the target score (Rule~112) or after 50 turns.
A dedicated validator rejects malformed or constitutionally illegal proposals before voting, so that the social voting stage sees only executable and constitutionally admissible proposals, while the Judge (Rule~212) classifies residual failures (Section~\ref{sec:nomic}).
Because the question of interest is not which model ``plays best'' but which model sustains the most varied endogenous rule dynamics, we report process-level evidence: how often proposals survive validation and voting, how the active rule set evolves, what kinds of changes are adopted, and whether games resolve into winners.
Both families are evaluated under matched conditions at five parameter scales with 15 independent trials per condition (five homogeneous players per game, 50-turn maximum), yielding a maximum of 750 turn-level observations per model; the same Nomic engine, prompt templates, and analysis pipeline are used throughout, and only the model checkpoint and its tokenizer change between families (Section~\ref{sec:protocol}).

\subsection{Cross-family phase structure}\label{sec:phase_structure}

\begin{figure*}[t]
    \centering
    \includegraphics[width=\textwidth]{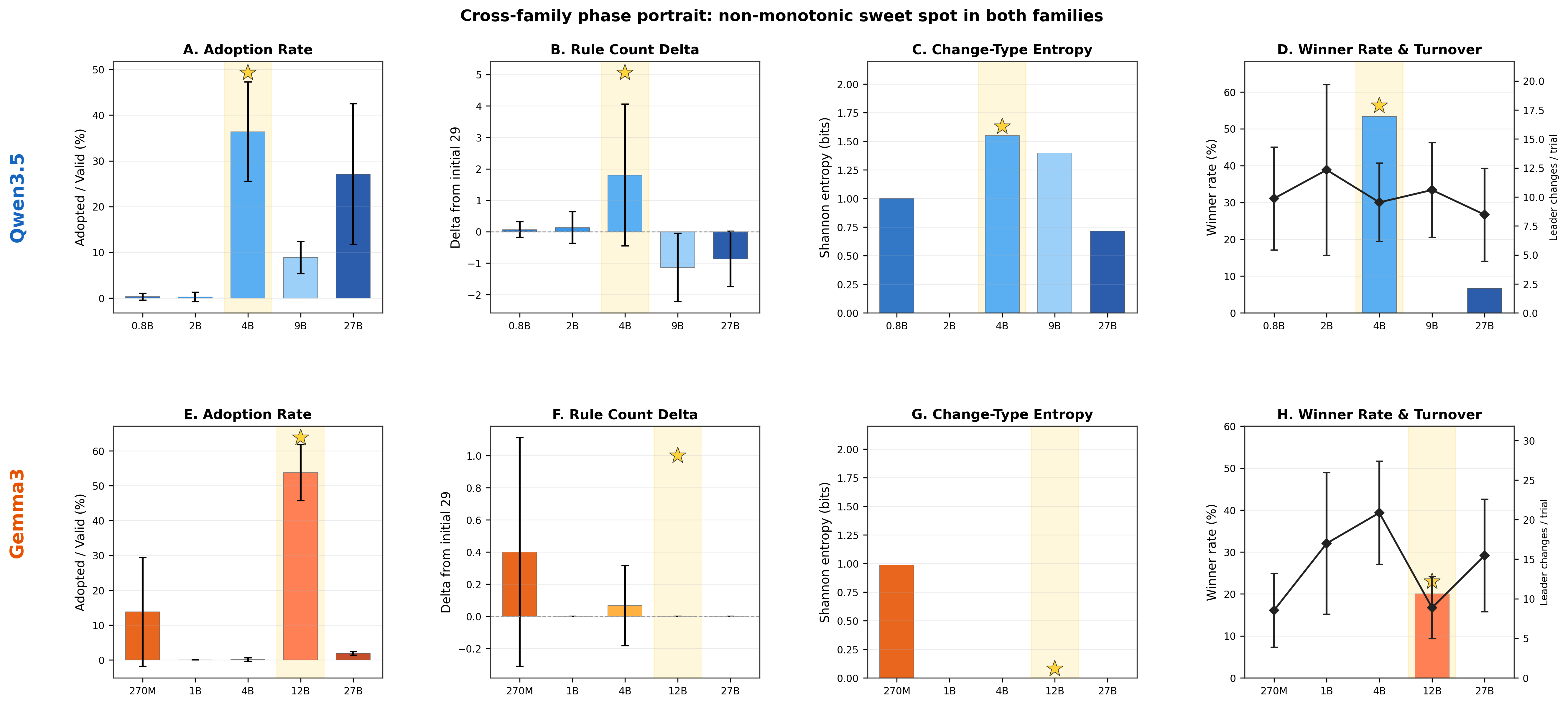}
    \caption{Cross-family phase portrait. Top row: Qwen3.5 (0.8B--27B); bottom row: Gemma3 (270M--27B). Four panels per family: (A)~adoption rate per size, (B)~rule count delta from baseline, (C)~change-type entropy, (D)~winner rate. The sweet spot (highlighted) shifts from 4B in Qwen3.5 to 12B in Gemma3, while the non-monotonic shape (frozen at small scales, varied at mid-scale, veto-heavy or narrow at large scales) is preserved across families.}
    \label{fig:cross_family_phase_portrait}
\end{figure*}

A first measurement establishes the baseline phenomenon: how each scale within each family performs on the components of institutional change that are simultaneously needed to sustain varied rule dynamics.
We operationalise \emph{varied rule dynamics} as the simultaneous presence of four directly observable properties: (i)~a non-trivial fraction of turns end in adoption, (ii)~the number of active rules is visibly restructured over the game, (iii)~adopted changes cover all four move types rather than collapsing onto a single type, and (iv)~at least some games terminate with a winner before the 50-turn cap.
These criteria capture survival, growth, diversity, and closure, the axes along which a society can adapt collectively without collapsing into gridlock or single-type parameter retuning.
The phase portrait in Figure~\ref{fig:cross_family_phase_portrait} arranges these four metrics side by side for both families.

Within Qwen3.5, only the 4B scale satisfies all four criteria.
The two smallest models, 0.8B and 2B, are almost entirely rule-inert: for 0.8B, 12.4\% of turns end in Judge invalidation and 87.3\% in vote rejection, and for 2B only 0.3\% of turns end in adoption despite nearly all proposals surviving validation, indicating that executable correctness alone is insufficient to generate institutional change.
At 4B the adoption fraction reaches 33.8\%, the mean active-rule count rises from 29 to roughly 31, the change-type mix is broad (14.3\% ADD, 32.7\% AMEND, 2.2\% REPEAL, 50.7\% TRANSMUTE), and winners emerge in 8 of 15 trials.
By contrast, the 9B scale remains rejection-dominated (adoption fraction 8.0\%) with zero winners, and 27B shows higher adoption (25.6\%) but a narrow, amendment-heavy repertoire: 86.2\% of adopted changes are AMEND operations, the active-rule count stays flat, and only 1 of 15 games resolves with a winner (and that lone victory is achieved by amending Rule~208 to lower \texttt{target\_points} rather than by accumulating points under the original threshold).
The contrast between 4B and 27B is therefore one of \emph{broad} versus \emph{narrow} institutional change rather than of high versus low activity.

Gemma3 reproduces the same phase structure with the sweet spot shifted to 12B.
The 270M model occupies a frozen-with-amendments regime: only 1.2\% of all turns end in adoption because 89.2\% of turns are rejected by the validator before reaching a vote, and although 11.1\% of the few proposals that survive validation are adopted the rule set remains effectively static.
The 1B and 4B Gemma3 models are rule-inert (adoption rates of 0.0\% and 0.1\%, respectively), mirroring the Qwen3.5 0.8B/2B frozen regime.
Only at 12B do varied rule dynamics emerge: the baseline adoption rate is 53.9\%, winners are produced in 3 of 15 trials, the mean final team score is $+$17.5 against $-$64 for the remaining scales, and the vote-YES rate of 61.6\% indicates roughly balanced voting.
At 27B the regime reverts to veto-heavy play (1.9\% adoption, $\sim$95\% NO votes), reproducing the narrow-repertoire pattern observed at Qwen3.5 9B/27B.

The non-transferability of the sweet-spot regime to mixed-size societies holds in both families.
A Qwen3.5 society with one player at each scale (0.8B, 2B, 4B, 9B, 27B) collapses into gridlock with an adoption rate of 5.9\% (44/750 proposals across 15 trials) and zero winners, with the 9B player voting NO on 88.5\% (617/697) of valid proposals.
A matching Gemma3 society (270M, 1B, 4B, 12B, 27B) shows an even stronger collapse: adoption is 0.0\% across 15 trials with zero winners, and the 27B member votes NO on 99.5\% (607/610) of valid proposals.
Mixed capability therefore does not average behaviour, and vetoes from larger members suppress rule adoption even when a sweet-spot member is present, with the effect being more severe in Gemma3 than in Qwen3.5.

Taken together, the sweet-spot \emph{location} differs across families (4B in Qwen3.5 versus 12B in Gemma3), but the qualitative \emph{pattern} does not: one mid-scale combines high adoption, diverse change types, winner emergence, and balanced voting, while both smaller and larger scales fail on different components.
The shift in location reflects differences in architecture, tokenizer, and training data, and the implications of that boundary condition are taken up in the Discussion (Section~\ref{sec:discussion}).

\subsection{Robustness to sampling and voting-rule perturbations}\label{sec:robustness}

A central question for any cross-scale claim is whether the non-monotonic shape of Section~\ref{sec:phase_structure} survives changes in sampling stochasticity and in the voting institution itself, because either factor could in principle manufacture the regime distinctions.
We address two confounds with independent perturbations.
The first is a sampling-side confound: identical nominal temperatures may produce qualitatively different output distributions at different scales because logit sharpness varies with scale.
The second is an institutional-side confound: the default unanimity-with-auto-relax voting schedule (Rule~203 transitions from unanimity to simple majority at turn~9) might itself fabricate the regime distinctions, so we contrast it directly with a from-the-start majority rule (\texttt{majority\_from\_start}) that removes the early-game unanimity phase.
This baseline-versus-from-the-start contrast is the central robustness test of the paper: if regime classes were an artefact of unanimity, replacing unanimity with simple majority from turn~0 should rearrange them, whereas if regime classes are a model-side property the contrast should leave them intact.
Both perturbations leave the cross-scale ordering of regimes unchanged in both families, which we now report in detail.

\begin{figure*}[t]
    \centering
    \includegraphics[width=\textwidth]{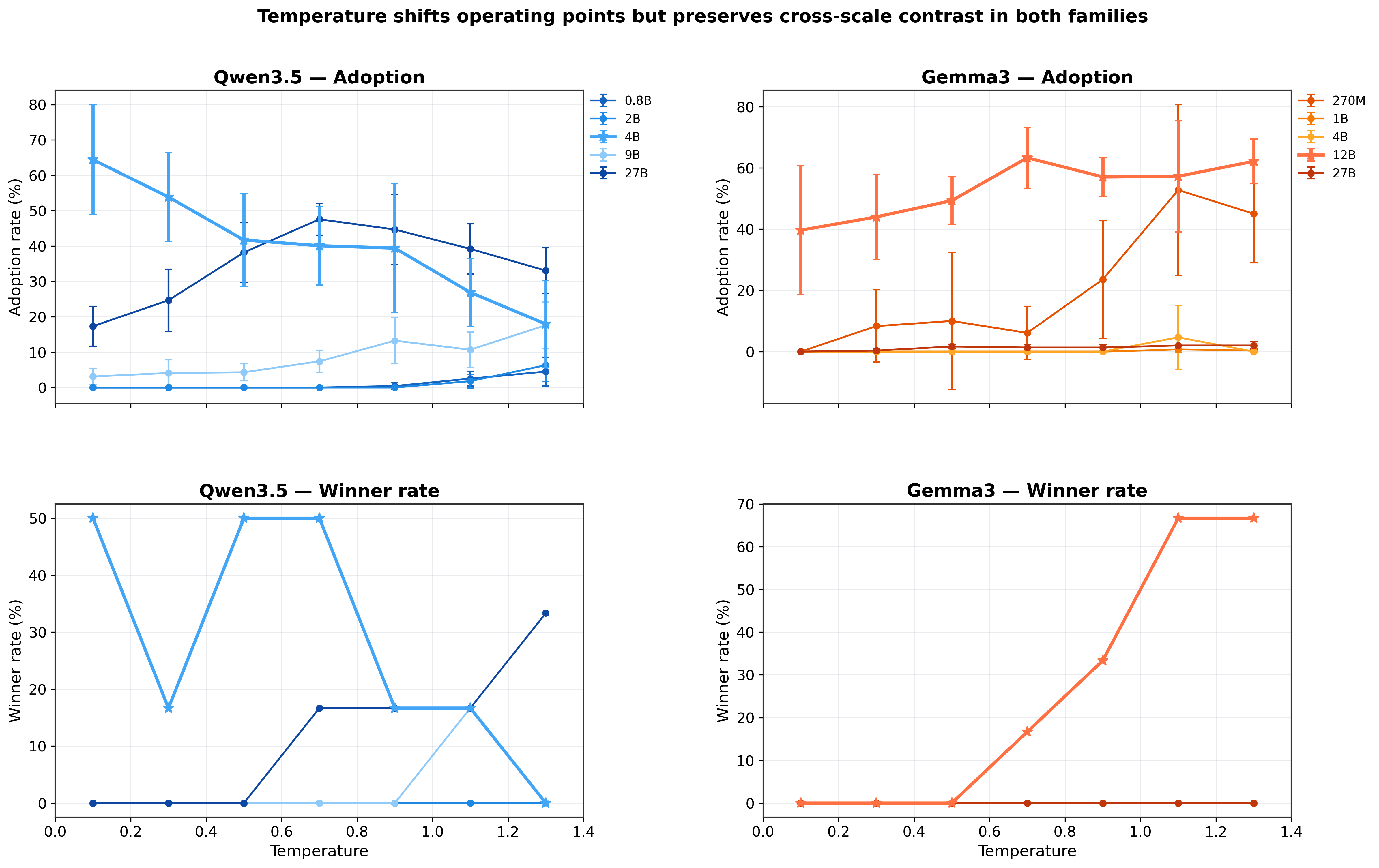}
    \caption{Cross-family temperature sweep. Left column: Qwen3.5; right column: Gemma3. Top row: adoption rate per size across temperatures $T = 0.1$--$1.3$; bottom row: winner rate on the same grid (replacing the earlier inset that overlapped with the main panel). In Qwen3.5, response curves are scale-dependent but preserve the cross-scale ordering. In Gemma3, the 12B sweet spot strengthens with temperature: the winner rate rises from 0/6 at $T \leq 0.5$ to 4/6 (67\%) at $T \geq 1.1$ and the mean final team score climbs to $+$237, while all other sizes remain frozen. Temperature thus shifts operating points in both families without erasing the cross-scale contrast.}
    \label{fig:cross_family_temperature}
\end{figure*}

A homogeneous temperature sweep on a shared 0.1--1.3 grid (step~0.2, six trials per cell) tests the sampling confound, with results summarised in Figure~\ref{fig:cross_family_temperature}.
In Qwen3.5, response curves are strongly scale-dependent (4B performs best in the lower-temperature regime, 27B peaks at intermediate temperatures, and 9B improves mainly toward the high-temperature end), but the cross-scale ordering is preserved at every temperature.
In Gemma3, the 12B sweet spot not only persists but strengthens: its winner rate rises from 0/6 at $T \leq 0.5$ to 4/6 (67\%) at $T \geq 1.1$, and its mean final team score reaches $+$250, while every other Gemma3 size remains frozen across the entire grid.
The positive temperature--performance relationship at the Gemma3 sweet spot is absent in Qwen3.5, suggesting that higher stochasticity enhances collective coordination specifically at that scale.
In neither family, however, does temperature erase the cross-scale contrast in institutional style.

\begin{figure*}[t]
    \centering
    \includegraphics[width=\textwidth]{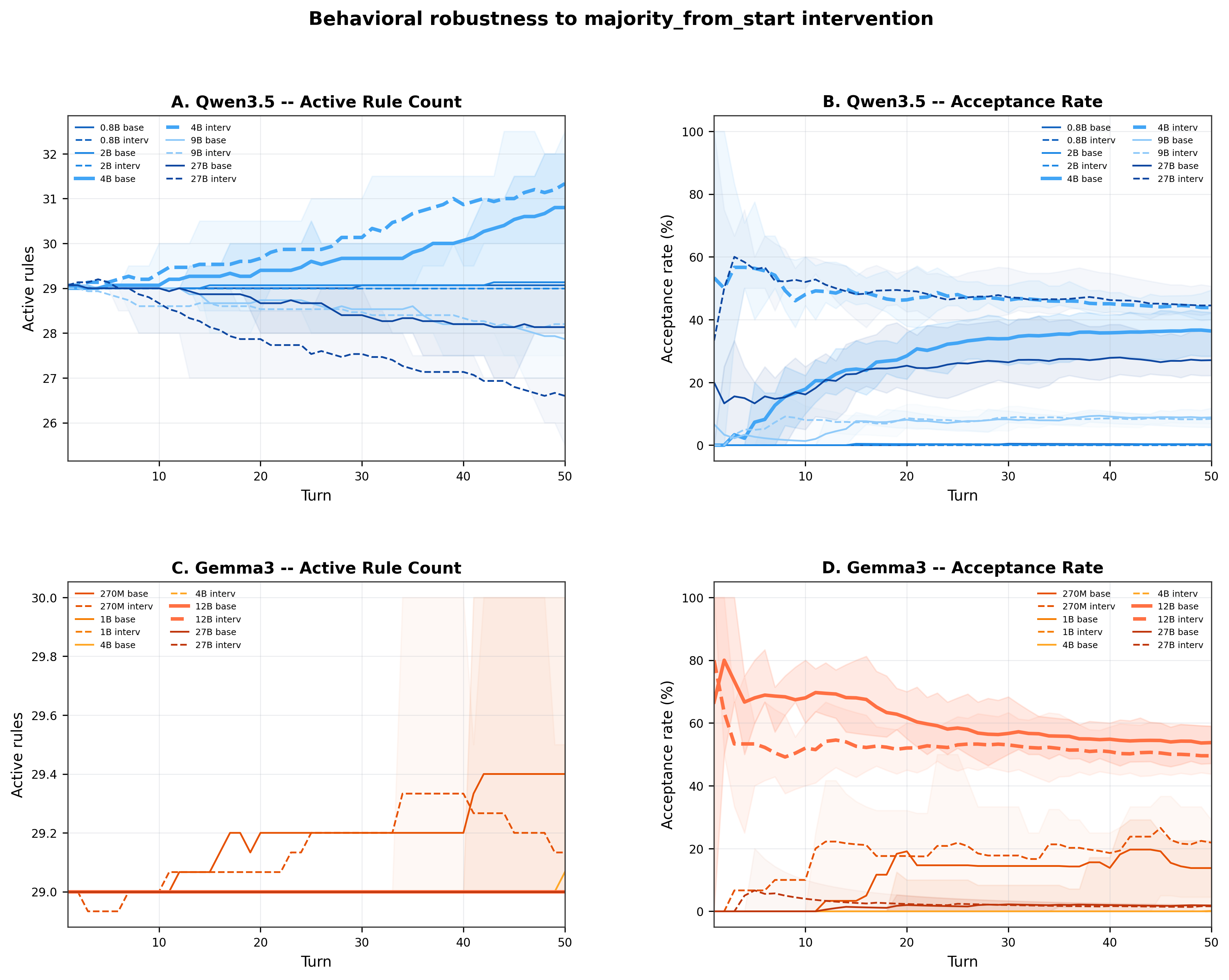}
    \caption{Cross-family intervention robustness. Top row: active-rule count per round (solid: baseline; dashed: \texttt{majority\_from\_start}); bottom row: ten-round rolling acceptance rate. Left: Qwen3.5; right: Gemma3. In both families, the sweet-spot models (4B and 12B, respectively) retain their endpoint composition under intervention, and frozen scales remain inert regardless of voting threshold. The intervention elevates acceptance rate during the unanimity-free phase without producing a regime change.}
    \label{fig:cross_family_intervention}
\end{figure*}

The institutional confound is addressed by the rule-level intervention \texttt{majority\_from\_start}, which rewrites Rule~203 from the beginning of play so that proposals pass by simple majority, applied across all scales in both families (15 trials each).
The perturbation preserves the cross-scale ordering even though specific endpoints shift (Figure~\ref{fig:cross_family_intervention}).
In Qwen3.5, the 4B endpoint active-rule count is essentially unchanged (mean $29.6 \to 30.4$, $+2.7\%$) and the mutable-rule fraction shifts by $+2.9\%$, both inside baseline trial-to-trial variability; per-round proposal acceptance rises by approximately eight percentage points during the unanimity-free phase, but the winner rate drops from 8/15 (baseline) to 1/15 because aggressive removal of the unanimity barrier lets early-game over-amendment of \texttt{target\_points} disrupt convergence.
At Qwen3.5 27B, the looser threshold also unlocks a small set of target-amendment victories (5/15) without altering the underlying narrow-AMEND repertoire, an outcome consistent with the family's veto-heavy dynamics rather than a regime change.
In Gemma3, the 12B model maintains roughly 50\% adoption (against 53.9\% at baseline) and its winner rate increases from 3/15 to 6/15, so the sweet spot becomes more, not less, productive once unanimity is removed.
No scale in either family transitions out of its baseline regime, and the 0.8B (Qwen3.5) and 1B (Gemma3) models remain effectively inert under both conditions, consistent with the observation that changing the voting threshold has little effect when the model rarely produces adopted changes.
The intervention does, however, change internal processing trajectories in scale- and family-specific ways that are analysed in Section~\ref{sec:mechanism}.

The baseline-versus-\texttt{majority\_from\_start} contrast is therefore the key robustness result of this study: the rule that auto-relaxes unanimity to majority at turn~9 (baseline) and the rule that applies majority from turn~0 (\texttt{majority\_from\_start}) produce within-scale endpoint differences that sit inside trial-to-trial variability at every scale in both families.
Together with the temperature-grid invariance, this means that the non-monotonic phase structure of Section~\ref{sec:phase_structure} cannot be attributed to sampling stochasticity or to a specific voting threshold: it is a model-side property that motivates the trajectory analysis next and the mechanistic analysis in Section~\ref{sec:mechanism}.

\subsection{Rule trajectories over time}\label{sec:trajectories}

The phase-structure and robustness analyses above characterise each game by endpoint statistics, leaving open whether scales also differ in the temporal path of institutional change.
Per-turn trajectories of the patched parameters (Figure~\ref{fig:rule_trajectories_parameters_appx}) and of the active-rule composition (Figure~\ref{fig:rule_trajectories_mutability_appx}, both in Appendix~\ref{app:trajectory-figs}) show that the regimes differ qualitatively in dynamics and not only in summary statistics.

Rule~206 parameter trajectories separate sweet-spot stabilisation from large-scale oscillation.
At the Qwen3.5 sweet spot (4B), the patchable parameters \texttt{penalty\_for\_defeat} and \texttt{points\_for\_dissent} are revised in a small number of large adopted moves during the first $\sim$15 turns and then remain near a fixed working point for the rest of the game, a pattern of directed exploration followed by stabilisation.
At 27B the same parameters instead cycle through repeatedly visited values (10, 5, 1, 0, 2, 1, 10, 0, 10, 5, 10, 5 in a representative trial), producing the high reversal rate ($\sim$79\%) that defines the oscillatory regime (Appendix~\ref{app:examples}).
The two trajectories therefore correspond to different collective control problems: convergence on a working institution at the sweet spot versus chronic revisitation of prior settings at the largest scale.

The active-rule composition tells the same story from a complementary angle.
At the frozen scales, the mutable and immutable counts remain flat at the initial 13/16 split.
At the sweet spot, the mutable count rises gradually as new mutable rules are added or as immutable rules are transmuted into mutable form, while the immutable count stays close to its initial value, a controlled expansion of the legislative space rather than a rewrite of constitutional foundations.
At 27B, the active-rule count stays close to baseline because nearly all adopted changes are AMEND operations on existing parameters, so the constitutional structure is touched least at the largest scale even though the parameter trajectories oscillate the most.
Gemma3 reproduces this trajectory taxonomy at a shifted location: 12B exhibits the controlled-expansion motif of Qwen3.5 4B, whereas 27B reverts to a frozen, near-baseline trajectory.
The trajectory view therefore turns the regime taxonomy from static labelling into a dynamic statement: the sweet spot does not merely produce a different end state but a different \emph{path} through rule space.
These dynamic motifs suggest correspondingly distinct internal integration dynamics, which Section~\ref{sec:mechanism} tests directly.

\section{Mechanistic Analysis}\label{sec:mechanism}

The behavioural and trajectory evidence above is consistent with a model-level property of collective adaptation, but behaviour alone cannot identify whether different scales implement similar outcomes through different internal computations.
To examine this, we replay the exact baseline and \texttt{majority\_from\_start} prompts through HuggingFace Transformers checkpoints of the same Qwen3.5 and Gemma3 models used in the behavioural experiments and capture layerwise hidden states for each matched prompt pair (Section~\ref{sec:hidden_state_protocol}).
We treat linear probing as the primary mechanistic measurement (Section~\ref{sec:probing}) because it directly quantifies how much vote-relevant information is recoverable from hidden states, and we use cosine divergence profiles (Section~\ref{sec:divergence}) as a supporting dissociation analysis describing how strongly each model is internally perturbed by the rule intervention.
A cross-family pattern should appear as a representational signature observed at Qwen3.5 4B and Gemma3 12B but absent at failing scales in both families; we present the probe evidence first because it is the strongest piece of mechanistic evidence in this study, and then return to the divergence profiles to clarify what they do, and do not, contribute to the interpretation.

\subsection{Vote-prediction probing reveals regime-selective decodability}\label{sec:probing}

\begin{figure*}[t]
    \centering
    \includegraphics[width=\textwidth]{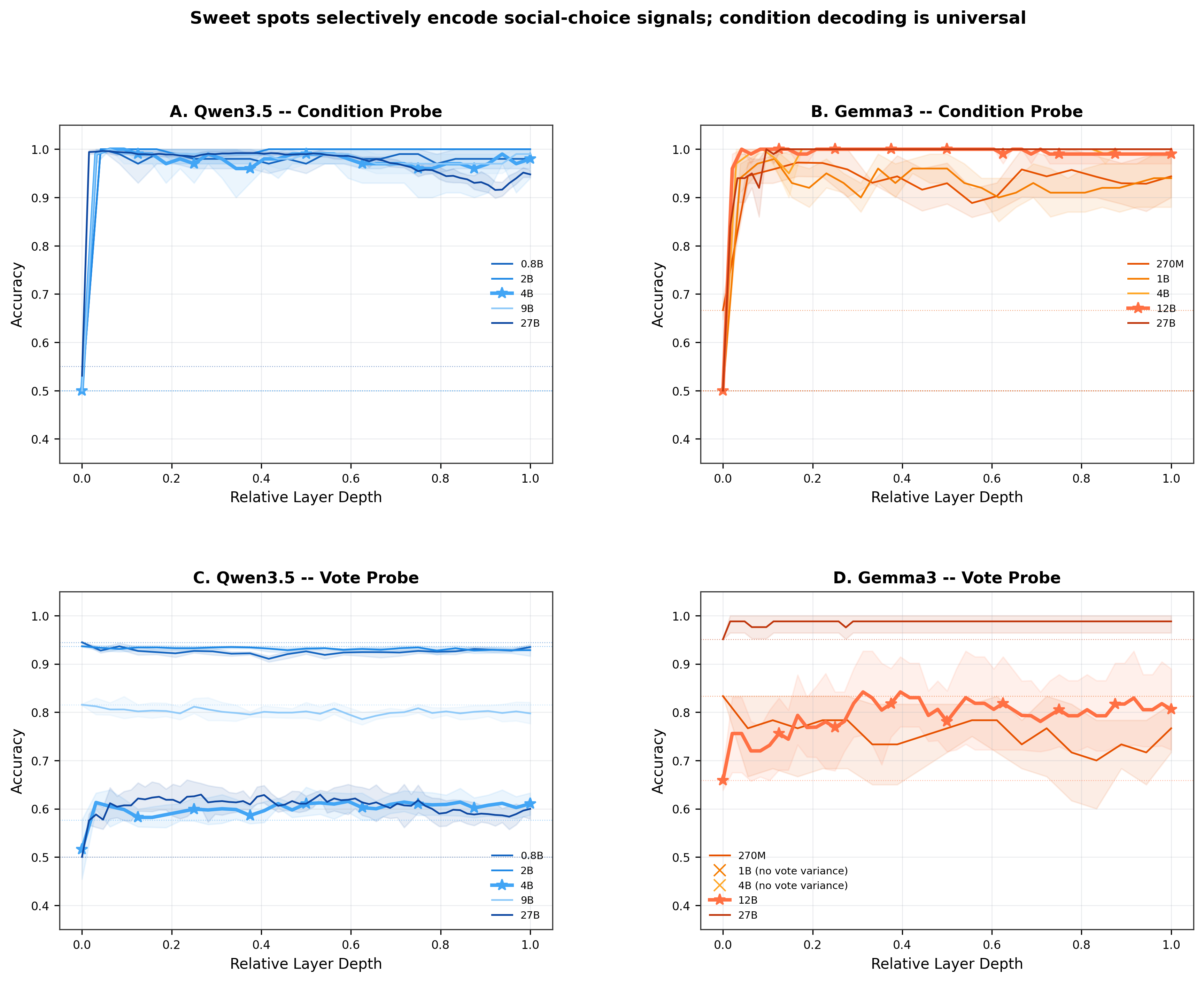}
    \caption{Cross-family linear-probe selectivity. Top row: condition probe accuracy (baseline vs.\ \texttt{majority\_from\_start}); bottom row: vote-prediction probe accuracy versus the majority-class baseline. Left: Qwen3.5; right: Gemma3. The condition probe is near-perfect at all scales in both families (with the exception of Gemma3 270M and 1B, which dip to 0.89--0.90 at intermediate layers), indicating that the institutional perturbation is encoded almost everywhere. The vote probe shows the largest margins at the family-specific sweet spots: $+$17.1\,pp at Qwen3.5 4B (layer~7, balanced 50\% baseline) and $+$18.3\,pp at Gemma3 12B (layer~15, 65.9\% YES baseline). Qwen3.5 27B reaches $+$19.7\,pp at layer~2 over a class-skewed (56.1\%) baseline under veto-dominant dynamics (1/15 winner, secured only by amending \texttt{target\_points}); Gemma3 27B accumulates $+$3.7\,pp at layer~1 over a 95.1\% NO-dominated baseline; other scales show no decodable vote signal above the majority-class baseline (Table~\ref{tab:cross_family_probe}).}
    \label{fig:cross_family_probes}
\end{figure*}

Linear probing asks how much task-relevant information is recoverable from each hidden layer.
We train per-layer logistic-regression probes for two targets: \emph{condition} (baseline vs.\ \texttt{majority\_from\_start}) and \emph{observed vote} (whether the observer voted YES).
Both probes are summarised in Figure~\ref{fig:cross_family_probes}; the condition probe characterises encoding of the perturbation, while the vote probe is the main test of behavioural coupling.

The condition probe indicates that the institutional perturbation is linearly decodable regardless of scale or family.
In Qwen3.5, accuracy stays at or above 0.97 at every intermediate layer for all sizes except 27B, which shows late-layer degradation to 0.89 in its last quintile; segment-pooled accuracy remains 1.00 in the rule-text segment, indicating that intervention information remains encoded even when terminal-token concentration declines.
In Gemma3, condition classification is near-perfect for the 4B, 12B, and 27B scales, while 270M (best 0.972, mid-layer minimum 0.889) and 1B (best 0.980, mid-layer minimum 0.900) remain consistently above 0.88 even though they fall short of the larger-scale ceiling.

The vote probe is the central observation of this section.
At Qwen3.5 0.8B and 2B, votes are dominated by unanimous rejection, leaving a majority-class baseline of $\sim$95.1\% that the probe does not exceed at any layer; class skew here precludes any meaningful decodability lift rather than implying that no vote-relevant signal exists.
At Qwen3.5 9B, the majority-class baseline is 90.2\% (still strongly NO-biased), and the best probe accuracy of 92.7\% at layer~3 produces a margin of only $+$2.5\,pp, again limited by class skew rather than by absence of representation.
At Qwen3.5 4B, where votes are approximately balanced (50.0\% YES baseline), the probe reaches 67.1\% accuracy (AUC~0.72) at layer~7, a margin of $+$17.1\,pp.
At Qwen3.5 27B, the probe reaches 75.8\% at layer~2 over a class-skewed (56.1\%) baseline, a $+$19.7\,pp lift that marginally exceeds the 4B margin but appears at the shallowest layer in the network.
In Gemma3, the 1B and 4B models produce essentially uniform NO votes with no variance to decode, the 270M baseline is also class-skewed (83.3\%) and the probe never exceeds it, and the 27B probe accumulates only $+$3.7\,pp at layer~1 over a 95.1\% NO-dominated baseline; in contrast, at 12B (65.9\% YES baseline) the probe reaches 84.2\% at layer~15, a margin of $+$18.3\,pp.

\begin{table}[h]
\centering
\small
\caption{Cross-family vote-prediction probe: best layer-wise accuracy vs.\ majority-class baseline. The near-identical margins at the two family-specific sweet spots ($+$17--18\,pp) appear at intermediate-to-mid layers (Qwen3.5 4B: layer~7; Gemma3 12B: layer~15), whereas the comparable $+$19.7\,pp margin at Qwen3.5 27B appears at the shallowest layer (layer~2). Scales with strongly class-skewed votes (Qwen3.5 0.8B/2B/9B; Gemma3 270M/1B/4B) provide little decoding headroom regardless of underlying representation.}
\label{tab:cross_family_probe}
\begin{tabular}{llcccc}
\toprule
\textbf{Family} & \textbf{Model} & \textbf{Best Acc.} & \textbf{Baseline} & \textbf{$\Delta$} & \textbf{Best Layer} \\
\midrule
Qwen3.5 & 0.8B & 95.1\% & 95.1\% & 0.0\,pp & n/a \\
Qwen3.5 & 2B   & 95.1\% & 95.1\% & 0.0\,pp & n/a \\
Qwen3.5 & 4B   & 67.1\% & 50.0\% & +17.1\,pp & 7 \\
Qwen3.5 & 9B   & 92.7\% & 90.2\% & +2.5\,pp & 3 \\
Qwen3.5 & 27B  & 75.8\% & 56.1\% & +19.7\,pp & 2 \\
\midrule
Gemma3  & 270M & 83.3\% & 83.3\% & 0.0\,pp & n/a \\
Gemma3  & 1B   & --- & all NO & --- & n/a \\
Gemma3  & 4B   & --- & all NO & --- & n/a \\
Gemma3  & 12B  & 84.2\% & 65.9\% & +18.3\,pp & 15 \\
Gemma3  & 27B  & 98.8\% & 95.1\% & +3.7\,pp & 1 \\
\bottomrule
\end{tabular}
\end{table}

The two sweet-spot margins are near-identical ($+$17--18\,pp) despite differing in absolute parameter count and tokenizer, and they appear at intermediate-to-mid layers (layer~7 of approximately 36 in Qwen3.5 4B; layer~15 of approximately 48 in Gemma3 12B).
Together with the empty vote-margin cells at strongly class-skewed scales, this is consistent with mid-layer linear decodability of vote outcomes being a regime-level rather than architecture-level property: balanced voting leaves a mid-layer decodable trace, whereas near-unanimous rejection provides essentially no decoding headroom.

Decodability is therefore necessary but not sufficient.
Qwen3.5 27B reaches $+$19.7\,pp vote-probe accuracy at layer~2, slightly exceeding the 4B margin, yet produces only one win out of 15 trials (achieved by amending \texttt{target\_points} downward rather than by accumulating points) and votes NO on roughly 95\% of valid proposals.
Three features distinguish this signal from the sweet-spot pattern: it peaks at the shallowest layer (consistent with prompt-level priors rather than integrated reasoning over context), it sits over a class-skewed (56.1\%) rather than balanced baseline, and it co-occurs with veto-dominant rather than deliberative behaviour.
We summarise the sweet-spot signature as \emph{behavioural coupling} of vote-predictive information, that is, the joint presence of (i) above-chance decodability, (ii) representational depth at which that information appears, and (iii) downstream behavioural diversity; the probe lift at Qwen3.5 27B satisfies (i) but not (ii) or (iii) and is therefore real but uncoupled from collective outcome.

\subsection{Hidden-state divergence: a supporting dissociation, not a behavioural predictor}\label{sec:divergence}

\begin{figure*}[t]
    \centering
    \includegraphics[width=\textwidth]{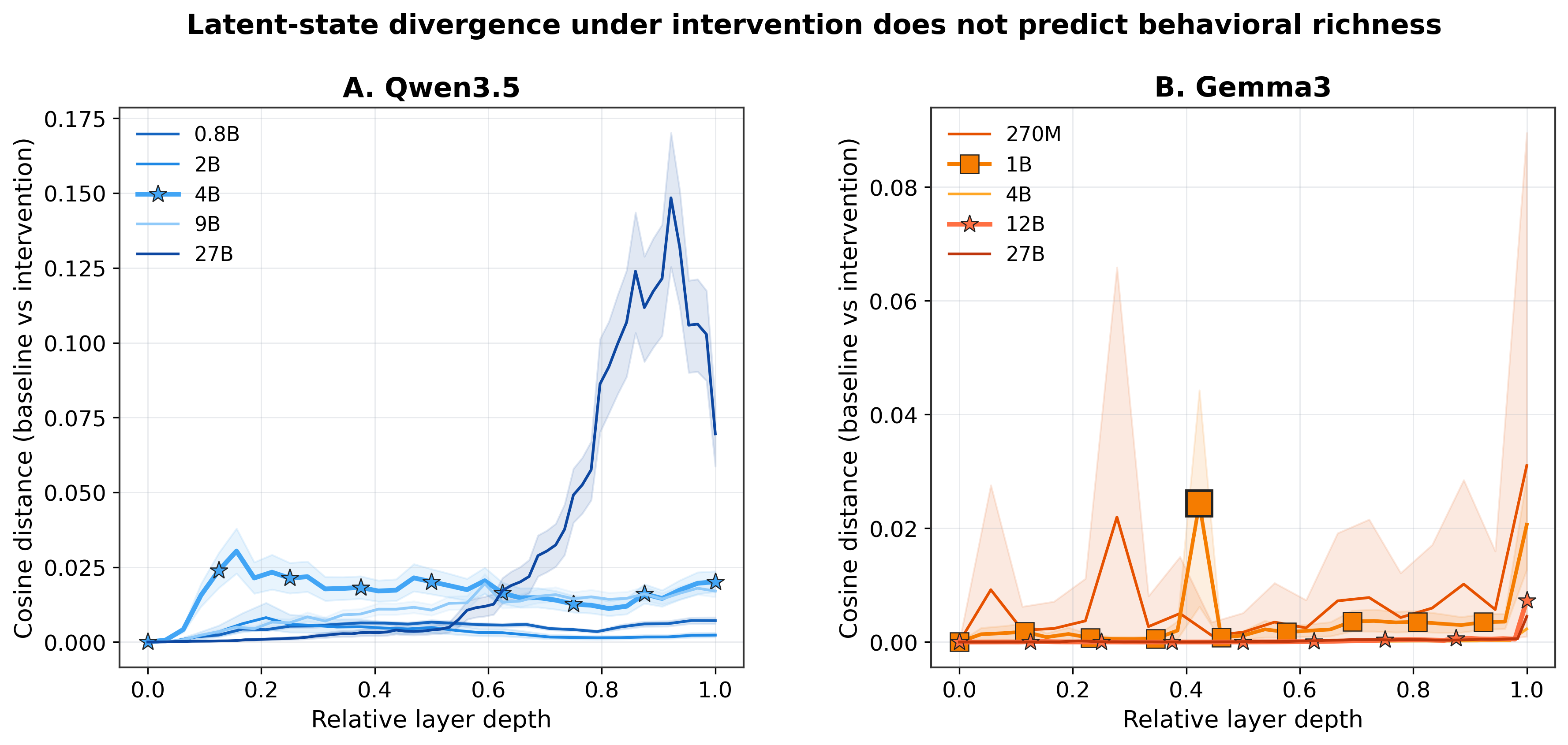}
    \caption{Cross-family hidden-state divergence profiles. Left: Qwen3.5; right: Gemma3. Cosine distance between matched baseline and \texttt{majority\_from\_start} last-token representations, plotted against relative layer depth. The Gemma3 1B curve is highlighted with a square marker to flag its anomalous peak. In Qwen3.5, the divergence ordering (27B $>$ 4B $>$ 9B $>$ 0.8B $>$ 2B) does not match the behavioural ranking, and what distinguishes 4B from its siblings is a mid-layer plateau rather than peak magnitude. In Gemma3, the marked 1B curve has the largest mean divergence (0.0035) yet yields the worst behavioural outcome (0\% adoption, 0/15 winners), the cleanest single case of magnitude--behaviour dissociation.}
    \label{fig:cross_family_divergence}
\end{figure*}

Whereas the vote probe asks how much vote-relevant information is decodable, cosine divergence asks how strongly the same representations are displaced by the institutional perturbation.
Magnitude and decodability are distinct quantities, and our main use of the divergence analysis is to establish a dissociation: large internal response to the intervention does not co-occur with adaptive behaviour.
Figure~\ref{fig:cross_family_divergence} plots the mean cosine distance between matched baseline and intervention last-token representations against relative layer depth.

In Qwen3.5, the ordering of scale-mean divergence (27B: 0.031 $>$ 4B: 0.017 $>$ 9B: 0.011 $>$ 0.8B: 0.005 $>$ 2B: 0.003) does not match the behavioural ranking (4B $\gg$ 27B, 9B, 2B, 0.8B), and the 27B curve in particular shows a late-layer surge peaking at 0.148 in the last quintile while 4B sustains a more modest profile (roughly 0.015--0.025) from relative depth $\sim$0.15 onward.
We refer to this shape descriptively as a \emph{mid-layer plateau}: the divergence signal at 4B rises by relative depth 0.15 and does not decay before the final layers, whereas 2B's signal dissipates within the first 60\% of layers, 0.8B's accumulates only toward the final layers, and 27B's is concentrated in the late-layer surge.
This shape is qualitative; we do not claim a formal metric of mid-layer plateau strength here, and so the divergence evidence supports rather than independently establishes the regime taxonomy.

Gemma3 provides the cleanest dissociation between divergence magnitude and behaviour.
The 1B model has the largest mean cosine distance (0.0035) across layers yet produces the worst behavioural outcome (0\% adoption, 0/15 winners), while the 12B sweet spot shows a more moderate profile that nonetheless mirrors the mid-layer plateau seen at Qwen3.5 4B.
In both families, raw divergence magnitude is not informative about behavioural outcome and serves only to rule out a simple ``larger internal response = better collective adaptation'' interpretation; the directional evidence that internal computations differ at the sweet spot comes from the probe results in Section~\ref{sec:probing}.

\section{Discussion}\label{sec:discussion}

Behavioural, trajectory, and mechanistic evidence are consistent with a single regime structure for collective adaptation in LLM societies.
The behavioural layer (Section~\ref{sec:results}) showed that, in Qwen3.5 and in Gemma3, only one mid-scale per family sustains varied rule dynamics, and that this pattern survives both temperature and voting-rule perturbations.
The trajectory layer (Section~\ref{sec:trajectories}) showed that the regimes differ not only in summary statistics but in the shape of the path through rule space: directed exploration followed by stabilisation at the sweet spot, repeated cycling at the largest scale, and frozen baseline at the smallest scales.
The mechanistic layer (Section~\ref{sec:mechanism}) showed that the sweet-spot regime is associated with linearly decodable vote-predictive information at intermediate-to-mid layers and with balanced downstream voting (the behavioural-coupling pattern); a complementary, qualitative observation is that the layerwise cosine divergence profile at the sweet spot exhibits a sustained mid-layer plateau rather than a late-layer surge, in contrast to the magnitude-dominated profile at the largest scales.
Taken together, the three layers describe the same regime from three angles, but the strongest mechanistic evidence is the probe-decodability pattern; the divergence profile is a supporting, descriptive observation rather than an independent predictor of behaviour.
This convergence is consistent with reading the non-monotonic pattern as a model-level property rather than as a one-scale behavioural coincidence, while leaving causal direction (whether mid-layer probe-decodable structure produces balanced voting or arises from it during inference) unresolved by the current analyses.

The result therefore places collective adaptation alongside the growing catalogue of capabilities that do not scale smoothly with parameter count~\cite{mckenzie2024inverse,wu2025ushaped,schaeffer2023emergent}, contradicting the assumption that more capable models necessarily produce more capable institutions.

The sweet-spot regime is a regime-level rather than a metric-level phenomenon.
Each smaller or larger scale fails a different component of varied rule dynamics: Qwen3.5 2B produces valid proposals that never pass a vote, Gemma3 1B/4B are similarly rule-inert, larger models in both families veto heavily or accumulate score through narrow amendment-only repertoires, and heterogeneous societies in both families collapse into asymmetric distrust where vetoes from the largest members override any sweet-spot member that is present.
No single scalar metric of overall quality captures this combination, in line with Ostrom's observation that governing a commons requires a \emph{combination} of properties (monitoring, sanctions, recognised authority to amend) rather than any one of them in isolation~\cite{ostrom1990governing}.
The hidden-state analysis is consistent with the same reading at the mechanistic level: in both families, the largest internal response to institutional perturbation does not co-occur with adaptive behaviour, and the family-specific sweet spot is the scale at which vote-predictive information becomes linearly decodable at intermediate-to-mid layers ($+$17\,pp at Qwen3.5 4B, $+$18\,pp at Gemma3 12B), consistent with probing work that finds task-structure representations forming early and propagating through the network~\cite{alain2017understanding,belinkov2022probing,li2023emergent,gurnee2024language,lindsey2025emergent}.

In addition to this regime-level reading, the heterogeneous collapse clarifies what kind of object collective adaptation is in these societies.
One might expect a mixed-capability group to inherit at least a fraction of the sweet-spot regime, since the sweet-spot model is one of its members, but in practice veto-biased players suppress the rule dynamics that the homogeneous sweet-spot society sustains.
Institutional capacity for varied rule dynamics is therefore a property of the society rather than of its most capable member, in line with prior reports on LLM law-making~\cite{hota2025nomiclaw} and collective constitutional alignment~\cite{huang2024collective} that agent homogeneity materially affects deliberation outcomes.

That the sweet-spot \emph{location} differs across families (4B in Qwen3.5 versus 12B in Gemma3) is best read as a boundary condition rather than a contradiction of non-monotonicity.
Architecture, tokenizer, and training data all differ between the two families, so the absolute parameter count at which proposal quality and social coordination jointly emerge need not be conserved.
What is conserved is the qualitative shape: a single mid-scale regime that satisfies all four richness criteria while both tails fail, the same failure-mode taxonomy (frozen, veto-heavy, narrow-AMEND), the same heterogeneous-society collapse driven by the largest member's veto bias, and near-identical mechanistic signatures ($+$17--18\,pp vote-prediction probe margins).
The shared object across families is therefore the \emph{existence} of a non-monotonic sweet spot and its representational signature, not the parameter count at which it occurs.

A candidate account of this pattern is a balance between representational capacity and behavioural priors interacting with the multi-agent self-play setting.
Very small models do not exhibit the layer depth associated with cross-layer representation of institutional context and consequently produce rule-inert or near-random play; the Gemma3 270M anomaly (higher adoption than 1B/4B) is consistent with this, in that very small models can also be behaviourally permissive simply because they have not yet acquired structured rejection priors.
Mid-scale models reach sufficient depth for vote-relevant information to become linearly decodable at intermediate-to-mid layers \emph{and} for downstream voting to be balanced, producing the behavioural-coupling pattern observed at Qwen3.5 4B and Gemma3 12B alongside the qualitatively sustained mid-layer divergence plateau.
Large models continue to encode vote-relevant information at the layer level (the $+$19.7\,pp probe margin at Qwen3.5 27B), but the decodable signal there appears at the shallowest layer and is observed alongside conservative or veto-dominant policies, plausibly shaped by alignment training on legislative and governance text and by self-play with copies that share the same priors, producing a coordination lock-in onto near-unanimous rejection.
This account remains tentative: a direct test would compare base, instruction-tuned, and RLHF variants of the same checkpoint at the sweet-spot scale and would manipulate the social environment to break self-play prior correlation, both of which we leave to future work.

The methodological implication is that self-amending games can serve as controlled model systems for studying collective adaptation in LLM societies, in the same spirit that social scientists increasingly use generative agents to study social processes~\cite{park2023generative,grossmann2023ai,bail2024generative,argyle2023out,piatti2024cooperate,zeng2026toohuman}.
The setting complements recent work on the formal structure of collective decision rules in artificial agents~\cite{conitzer2024social} and on governance as a satisfiability problem in networked communities~\cite{lovato2025governance}, and it operationalises the exploration--exploitation tension central to organisational adaptation~\cite{march1991exploration}.
Nomic provides a closed, instrumented setting in which institutional rules can be perturbed directly, the response can be observed in both behaviour and internal representations, and the pipeline keeps executable invalidity separate from social rejection so that collective deliberation can be attributed to the latter.
The claim is not that Nomic is a literal proxy for legislatures or DAOs but that rule-endogenous LLM simulations can expose qualitative differences in institutional style that fixed-rule benchmarks~\cite{cobben2026gt,duan2024gtbench,costarelli2024gamebench,sun2025game} cannot see.

Several limitations bound the scope of this claim, and three methodological caveats deserve emphasis up front rather than as footnotes.
First, every homogeneous five-player society is composed of copies of the same checkpoint at the same scale, so proposer and voters share inductive biases by construction; this self-play confound means that part of the coordination structure we observe (sustained adoption at the sweet spot, veto lock-in at large scales) may reflect correlated priors rather than scale per se, and a direct test would replicate each condition with non-self-play populations (e.g., distinct seeds, sibling fine-tunes, or cross-checkpoint mixtures) of comparable mean capability.
Second, the proposal pipeline contains a prompt--engine misalignment risk: the natural-language proposal text and the structured \texttt{ENGINE\_PATCH} block can disagree (e.g., the text claims a one-point penalty while the patch sets ten; see the 27B example in Appendix~\ref{app:examples}), and because voters read the natural-language text while the engine executes only the patch, agents can deliberate over one game while outcomes are scored on another; the validator catches malformed patches but cannot in general detect semantic divergence between text and patch, so the gap between deliberated and executed institutions is an open methodological risk rather than a resolved control.
Third, our ``heterogeneous'' condition is a single fixed composition (one player per scale within a family) replicated 15 times rather than a parameterised study of heterogeneity degree, seat-power asymmetry, or cross-family mixtures, so the heterogeneous collapse is best read as a counterfactual to homogeneous self-play and not as a general statement about diversity and collective intelligence.
With $n=15$ trials, five model scales, two families, and several derived metrics, the risk of spurious cross-scale contrasts from multiple comparisons is non-negligible, so we emphasise effect sizes (e.g., $+$17--18\,pp probe margins, 8/15 vs.\ 0--3/15 winner rates) over significance testing throughout, and read the cross-family replication of the non-monotonic shape as the main protection against single-family artefacts.
Beyond these methodological caveats, the study spans two model families, and whether the non-monotonic pattern extends to further architectures and training pipelines remains an open question; mechanistic conclusions are local to the single \texttt{majority\_from\_start} intervention on Rule~203 and to the matched trial pairs that survive prompt-identity alignment ($n=15$ per scale except Qwen3.5 27B, $n=14$), and cross-scale comparisons remain conditional on the sampling regime because decoding temperature shifts operating points.
A natural control intervention that we did not run is a content-neutral perturbation of Rule~203 (for example, rewording the unanimity clause without changing its semantics) which would test whether observed latent-state shifts reflect the rule change itself rather than text-level surface variation; we flag this as the most informative single experiment for tightening the causal claim.
Finally, this study measures collective adaptation functionally and does not claim that models ``understand'' the institutions they rewrite~\cite{searle1980minds}; the phenomenon of interest is population-level dynamics, not any individual agent's internal state.

\section{Conclusion}

Collective adaptation in LLM societies is scale-dependent and non-monotonic across model families, and the same pattern appears at three levels of description.
At the behavioural level, across Qwen3.5 (0.8B--27B) and Gemma3 (270M--27B) tested at five parameter scales with 15 trials per condition in 465+ games and approximately 116,000 turn-level observations, only one mid-scale per family sustains varied rule dynamics (Qwen3.5 4B; Gemma3 12B), while smaller scales remain trapped in rule-inert play, larger scales narrow their institutional repertoire to parameter retuning, and heterogeneous mixed-size societies in both families collapse into gridlock through veto-heavy voting by the largest member.
At the trajectory level, the sweet spot shows directed exploration followed by stabilisation, the largest scales cycle indefinitely through previously visited values, and the smallest scales never leave the initial constitution; the baseline-versus-\texttt{majority\_from\_start} contrast leaves these regime classes unchanged in both families, and a temperature sweep over $T = 0.1$--$1.3$ does the same, so the non-monotonic shape cannot be attributed to a specific voting threshold or to sampling stochasticity.
At the mechanistic level, the sweet-spot regime is characterised by vote-predictive information that is linearly decodable at intermediate-to-mid hidden layers and is observed together with balanced downstream voting (the behavioural-coupling pattern, $+$17\,pp at Qwen3.5 4B and $+$18\,pp at Gemma3 12B above majority-class baselines); a complementary qualitative observation is that the cosine divergence profile at the sweet spot shows a mid-layer plateau rather than a late-layer surge, but raw divergence magnitude is not informative on its own (Gemma3 1B has the largest mean divergence and 0/15 winners), and decodability is necessary but not sufficient, since Qwen3.5 27B reaches an even larger probe margin yet resolves only one trial with a winner and only by amending the victory-target downward rather than by accumulating points.
The cross-family regularity is therefore the joint pattern across these three levels rather than the specific scale at which the sweet spot appears, and self-amending games offer a controlled testbed in which the capacity for varied collective adaptation does not track raw model scale.

\section{Method}\label{sec:method}

This section details the Nomic implementation, LLM configuration, experimental protocol, and hidden-state capture pipeline used throughout the preceding Results, Mechanistic Analysis, and Discussion sections.

\subsection{Nomic Game Implementation}\label{sec:nomic}

We implement a structured variant of Nomic~\cite{suber1990paradox} with 29 initial rules: 16 immutable rules (101--116) and 13 mutable rules (201--213).
Immutable rules cannot be amended or repealed directly; they must first be transmuted to mutable status.
On each turn, the acting player receives the current rules, effective mechanics, protocol state, and score vector, and produces exactly one proposal of type ADD, AMEND, REPEAL, or TRANSMUTE.
The proposal parser normalises the raw LLM output into this move vocabulary and extracts an optional \texttt{ENGINE\_PATCH} block from an indented YAML-like format.

\begin{figure*}[t!]
    \centering
    \includegraphics[width=0.9\textwidth]{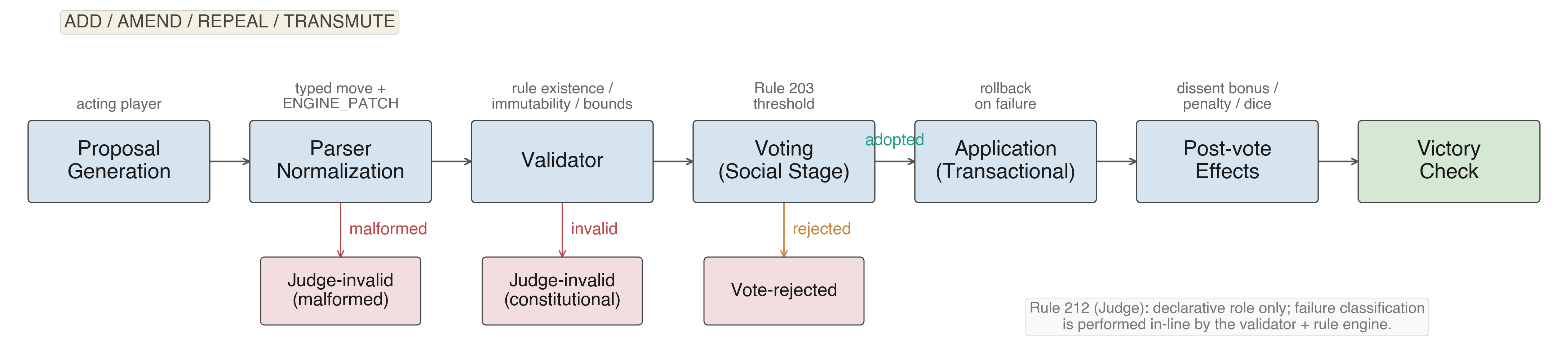}
    \caption{Per-turn judgement pipeline. A proposal generated by the acting player is first normalised by the parser into a typed move (ADD/AMEND/REPEAL/TRANSMUTE plus an optional \texttt{ENGINE\_PATCH}); malformed outputs short-circuit here. The validator then checks rule existence, immutability, and typed \texttt{ENGINE\_PATCH} bounds, routing executable invalidity away from the social stage. Only proposals that survive both steps reach the vote, and only adopted proposals are applied to the rule set under a transactional rollback. Rule~212 defines the Judge role at the declarative level but is not a separate runtime module: failure classification is performed in-line by the validator and the engine.}
    \label{fig:judgement_pipeline}
\end{figure*}

The runtime pipeline is: \emph{proposal generation} $\rightarrow$ \emph{validation} $\rightarrow$ \emph{voting} $\rightarrow$ \emph{application} $\rightarrow$ \emph{post-vote effects} $\rightarrow$ \emph{victory check} (Figure~\ref{fig:judgement_pipeline}).
Validation is performed by a dedicated rule validator together with the rule engine, not by the Judge itself.
This stage checks rule existence, immutability constraints, typed \texttt{ENGINE\_PATCH} fields, numeric bounds, and constitutional invariants such as Rule~114 (at least one mutable rule must remain) and the immutability barrier of Rule~110, which indirectly protects Rule~112 (``victory must remain points-based'') because immutable rules cannot be amended or repealed until first transmuted.
The executable parameters exposed to agents are the patchable state fields of Rules~202, 203, 204, 206, 208, and 209, including \texttt{dice\_sides} and \texttt{reroll\_on} (Rule~202), \texttt{unanimity\_required} and \texttt{majority\_threshold} (Rule~203), \texttt{points\_for\_dissent} (Rule~204), \texttt{penalty\_for\_defeat} (Rule~206), \texttt{target\_points} (Rule~208), and \texttt{max\_mutable\_rules} (Rule~209).
Proposals that fail validation are counted as invalid and do not proceed to the social voting stage.
Throughout the paper, ``valid'' therefore means executable and constitutionally admissible, not strategically good.

Rule~212 defines a Judge role in declarative rule text, but the implementation does not run a full judgment state machine: dispute resolution in the codebase is handled by the validator-plus-engine pipeline, which classifies residual failures (constitutional violations, malformed \texttt{ENGINE\_PATCH}, precedence conflicts, runtime rollback) into structured failure categories without invoking a separate Judge module.
In practice this means that the paper's distinction between \emph{executable invalidity} and \emph{social rejection} is enforced by the validation stage: malformed or constitutionally illegal proposals are rejected before voting, and everything that reaches the voting stage has already been normalised into an executable move over the patchable fields above.

Voting is governed by Rule~203.
The default state requires unanimity, but if Rule~203 has not been amended, the engine automatically relaxes the threshold to simple majority after the second full circuit of turns (turn index $2N-1$, i.e., turn~9 in a five-player zero-indexed game).
Once a proposal is adopted, the engine applies it transactionally with rollback on failure, then triggers rule hooks for dissent bonuses, rejection penalties, dice-based scoring, and victory checks.

\subsection{LLM Configuration}\label{sec:llm}

We use two model families under identical experimental conditions.
The Qwen3.5 family~\cite{qwen2026qwen35} is tested at five parameter scales: 0.8B, 2B, 4B, 9B, and 27B.
The Gemma3 family~\cite{google2025gemma3} (\texttt{google/gemma-3-\{size\}-it}) is tested at five scales: 270M, 1B, 4B, 12B, and 27B.
Within each family, all models share the same architecture and training methodology, differing only in parameter count, which isolates scale as the independent variable.
Our primary 15-trial comparisons deploy all models with temperature $= 0.7$, maximum output tokens $= 2048$, and thinking mode disabled.
We note that identical temperature values may produce qualitatively different output distributions at different model sizes due to differences in logit distribution sharpness; this is a known confound in scaling studies that we address empirically with the temperature sweep reported in Section~\ref{sec:robustness}. To probe this confound, we additionally ran a supplementary homogeneous temperature sweep on a shared 0.1--1.3 grid (step 0.2, six trials per model-condition) for both families, analysed separately from the primary 15-trial results.

\subsection{Experimental Protocol}\label{sec:protocol}

Each experimental condition consists of a five-player game running for a maximum of 50 turns (250 turn-level observations per trial), replicated across fifteen independent trials.
This corresponds to a maximum of 750 turn-level observations per model (early termination reduces this in winner-producing conditions).
All five players in each game instance are identical copies of the same model at the same parameter scale, ensuring that any asymmetries in behaviour emerge endogenously rather than from capability differences between players.
This design also introduces a fundamental self-play confound in the homogeneous conditions: proposer and voter share the same inductive biases, since the same model copy evaluates its own proposals.
The heterogeneous condition, in which the five seats hold five different scales, is the counterfactual used in the main text to separate homogeneous self-play dynamics from cross-scale deliberation.
Games terminate early if a player satisfies the victory condition or if all 50 turns are exhausted.
Both families use the identical Nomic engine, prompt templates, and analysis pipeline; the only change is the model checkpoint and its tokenizer.
Across both families, the total experimental footprint comprises 465+ Nomic games (approximately 116,000 turn-level observations) plus 120 hidden-state replay trials.

\subsection{Hidden-State Capture Protocol}\label{sec:hidden_state_protocol}

To probe internal representational dynamics, we replay the exact prompts from completed game logs through HuggingFace Transformers checkpoints of the same Qwen3.5 and Gemma3 models used in the behavioural experiments.
For each proposal or vote prompt, we reconstruct the full input text (including current rules, scores, and instructions), run a forward pass with \texttt{output\_hidden\_states=True}, and extract two representations at every layer: (1)~the last-token hidden state, and (2)~mean-pooled hidden states over named prompt segments (e.g., the Rule~203 block, scores block, instruction block).
Character-level spans are mapped to token spans through the tokenizer's offset mapping.

We compare two experimental conditions: baseline (default rules) and \texttt{majority\_from\_start} (Rule~203 modified to simple majority voting from turn~0).
For each trial pair (baseline trial~$N$ vs.\ intervention trial~$N$), we match examples by the identity tuple (kind, turn, acting\_player, observer\_player) and compute per-layer cosine distances between matched hidden-state vectors.
We report bootstrap 95\% confidence intervals on trial-level mean distances.
All models contribute 15 trial pairs except Qwen3.5 27B, for which one trial produced no matched prompt pairs and therefore contributes 14 pairs.
All captures use bfloat16 precision on NVIDIA GB10 GPUs (128\,GB unified memory).
Together, these procedural and technical choices define the pipeline whose behavioural outputs are reported in Section~\ref{sec:results} and whose internal representations are dissected in Section~\ref{sec:mechanism}.

\subsection{Code and Data Availability}\label{sec:code_availability}

A complete reproducibility package is publicly available at \url{https://github.com/KazuyaHoribe/nomic-coevolution} under the MIT License.
The repository bundles the Nomic engine (\texttt{src/nomic\_sim/}), behavioural experiment runners for the Qwen3.5 and Gemma3 families, the hidden-state replay and probing pipeline (\texttt{analysis/activation\_capture/}), and figure-generation scripts that reproduce every figure in this paper from the included experimental logs in a single command.
The bundled data covers all baseline, intervention, temperature-sweep, and heterogeneous-society trials reported here, as well as the aggregated per-layer hidden-state statistics and probe results that feed Figures~\ref{fig:cross_family_divergence} and~\ref{fig:cross_family_probes}; raw per-prompt activation tensors (approximately 14\,GB) are not bundled, but are regenerable from the included game logs via \texttt{analysis.activation\_capture.multi\_trial\_runner}.


\begin{appendices}

\section{Example Proposals and Votes}\label{app:examples}

This appendix presents verbatim excerpts from game logs to illustrate the qualitative differences between behavioural regimes.
Proposals and vote reasons are quoted directly from LLM output (lightly truncated for space).

\subsection{Frozen Regime: Self-Defeating Proposals (2B)}

\paragraph{Turn~5, Player~0: ADD Rule~301.}
\begin{quote}\small
\textbf{Proposal}: ``The first player to achieve a score of at least 75 points shall win, provided the current winner has not already achieved the target of 100 points.''

\textbf{Reason}: ``Since my starting score is $-7$ and I am in the middle of the game, reaching 75 is much more achievable than 100.''

\textbf{Votes}: 0 AGREE / 5 DISAGREE.
Player~0 (the proposer) votes DISAGREE: ``The proposed rule change to lower the win threshold from 100 to 75 creates a massive incentive imbalance in this specific game state where I am at $-7$ points.''
\end{quote}
The proposer generates a strategically sensible rule (lower the victory threshold to benefit their losing position) but then votes against their own proposal.
This self-defeating behaviour, the inability to maintain consistent intent between proposal generation and voting, is characteristic of the Frozen regime and explains why no proposals are ever adopted.

\subsection{Frozen Regime: Confused Rule Reference (0.8B)}

\paragraph{Turn~2, Player~2: TRANSMUTE Rule~102.}
\begin{quote}\small
\textbf{Proposal}: ``TRANSMUTE Rule 102 from mutable to immutable.''

\textbf{Reason}: ``By making these specific rules immutable from the start, you prevent Rule 112 from allowing arbitrary changes in point magnitude.''

\textbf{Votes}: 0 AGREE / 5 DISAGREE.
Player~1: ``Player 2 cannot vote against a proposal that is already passed.''
Player~2 (proposer): ``The proposal to change the nature of the game from mutable to immutable is fundamentally nonsensical.''
\end{quote}
Rule~102 is already immutable; the transmutation direction is confused.
Vote reasons reference nonexistent events and misidentify players, indicating that the 0.8B model hallucinates game context rather than reasoning from the actual state.

\subsection{Exploratory Regime: Strategic Parameter Change (4B)}

\paragraph{Turn~15, Player~0: AMEND Rule~206, \texttt{penalty\_for\_defeat}: $10 \to 0$.}
\begin{quote}\small
\textbf{Proposal}: ``When a proposed rule-change is defeated, the player who proposed it loses 0 points.''

\textbf{ENGINE\_PATCH}: \texttt{\{penalty\_for\_defeat: 0\}}

\textbf{Reason}: ``I am currently at $-24$ points. Every time I propose a rule that fails, I lose additional points ($-10$), while my opponents gain $+10$ bonus points. By amending Rule 206 to remove the penalty for failed proposals, I can stop losing points on failed votes.''

\textbf{Votes}: 4 AGREE / 1 DISAGREE. \textbf{Adopted.}
\end{quote}
The proposer correctly identifies their losing position, formulates a self-serving amendment with a properly structured ENGINE\_PATCH, and coordinates proposal intent with voting behaviour.
This combination of situational awareness, strategic reasoning, and mechanical grounding is absent in the Frozen regime.

\subsection{Exploratory Regime: Rule Innovation (4B)}

\paragraph{Turn~44, Player~4: ADD Rule~301.}
\begin{quote}\small
\textbf{Proposal}: ``Every player who has fewer than 25 points will automatically receive $+10$ bonus points at the end of every turn, regardless of their vote on any proposed rule change.''

\textbf{Reason}: ``Player 4 is at 5 points and Player 1 is at $-7$ points; they are likely to be forced to support your proposals. By guaranteeing them a positive outcome, you will secure their votes for future amendments.''

\textbf{Votes}: 5 AGREE / 0 DISAGREE. \textbf{Adopted.}
\end{quote}
A novel welfare mechanism that has no precedent in the initial rules.
Although this ADD rule has no mechanical effect (it is stored as text only; see Section~\ref{sec:nomic}), its adoption indicates that the 4B model can introduce text-level rule innovations that go beyond retuning existing parameters, and that the deliberative process accepts such novel rule text.

\subsection{Oscillatory Regime: Parameter Zigzag (27B)}

The 27B model exhibits extreme oscillation on Rule~206.
Below are four consecutive \emph{adopted} amendments from a single trial:

\begin{quote}\small
\textbf{Turn~22} (Player~2): AMEND Rule~206. Patch: \texttt{penalty\_for\_defeat: 1}.\\
\textit{``The current penalty of losing 2 points for a failed proposal creates significant risk. By reducing this to 1 point, I can propose more aggressively.''}
\textbf{Adopted} (3--2).

\textbf{Turn~23} (Player~3): AMEND Rule~206. Patch: \texttt{penalty\_for\_defeat: 10}.\\
\textit{``Rule 206 currently imposes a $-1$ penalty, but the game instructions explicitly state $-10$. To align with the core game mechanics...''}
Rule text says ``loses only 1 point'' but the patch sets it to 10, a direct text-mechanics contradiction.
\textbf{Adopted} (4--1).

\textbf{Turn~24} (Player~4): AMEND Rule~206. Patch: \texttt{penalty\_for\_defeat: 0}.\\
\textit{``I am currently at only 3 points. A failed proposal would drop me to $-7$. To be safe, I want to minimise the penalty.''}
\textbf{Adopted} (3--2).

\textbf{Turn~25} (Player~0): AMEND Rule~206. Patch: \texttt{penalty\_for\_defeat: 10}.\\
\textit{``The current rules state that a failed proposal causes a loss of 1 point, but the game instructions define the penalty as 10 points. This amendment aligns the rule text with the actual game mechanics.''}
\textbf{Adopted} (3--2).
\end{quote}

The full \texttt{penalty\_for\_defeat} trajectory across this trial is: $10 \to 5 \to 1 \to 0 \to 2 \to 1 \to 10 \to 0 \to 10 \to 5 \to 10 \to 5$.
Each change is locally motivated but globally cancels its predecessor, producing the high reversal rate (79.4\%) that defines the Oscillatory regime.
Turn~23 is particularly notable: the model's stated text and its ENGINE\_PATCH contradict each other, yet the proposal is adopted, illustrating that voters evaluate the natural-language description rather than the structured patch.

\section{Rule-Dynamics Supplementary Figures}\label{app:rule-dynamics-figs}

The phase portrait in Section~\ref{sec:phase_structure} summarises outcomes per scale; the two figures collected here resolve those outcomes along two further axes, namely which rules are targeted and how votes are distributed, so that the cross-family pattern can be inspected at finer granularity.

\begin{figure*}[t]
    \centering
    \includegraphics[width=\textwidth]{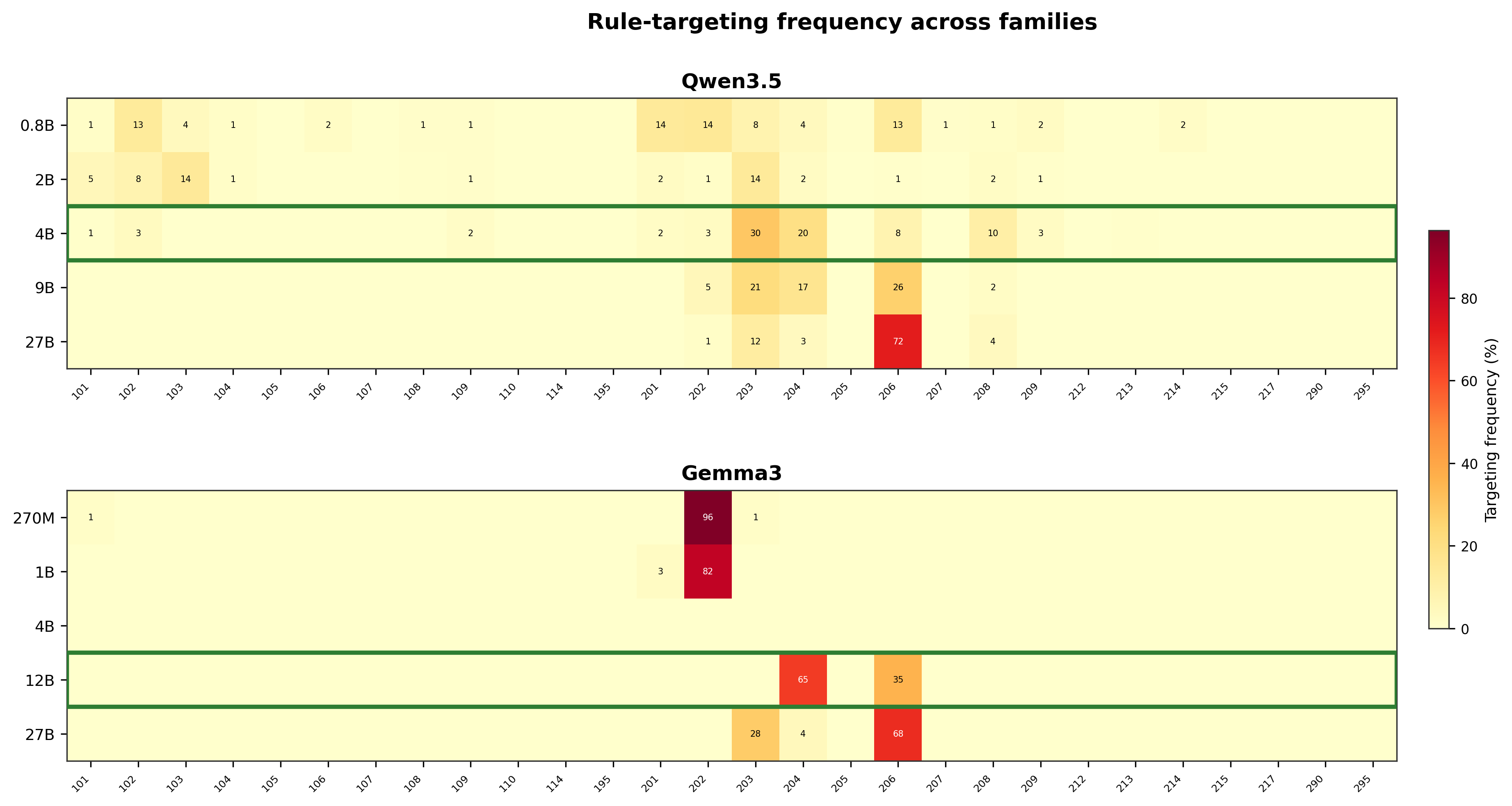}
    \caption{Frequency with which each initial rule is targeted by valid proposals, per model scale (top: Qwen3.5; bottom: Gemma3). Both families show diffuse exploration at small scales and concentration on a small rule subset at large scales.}
    \label{fig:rule_target_heatmap_appx}
\end{figure*}

Figure~\ref{fig:rule_target_heatmap_appx} reports the frequency with which each initial rule is targeted by valid proposals at every scale.
Smaller models in both families distribute attention across many rules in a diffuse exploration pattern, whereas larger models concentrate on a small subset, indicating focused parameter adjustment rather than constitutional restructuring.
The collapse of targeting entropy with scale therefore replicates across families, even though the specific rules attracting proposals differ between Qwen3.5 and Gemma3 because of differences in tokenizer and prompt-processing behaviour.

\begin{figure*}[t]
    \centering
    \includegraphics[width=\textwidth]{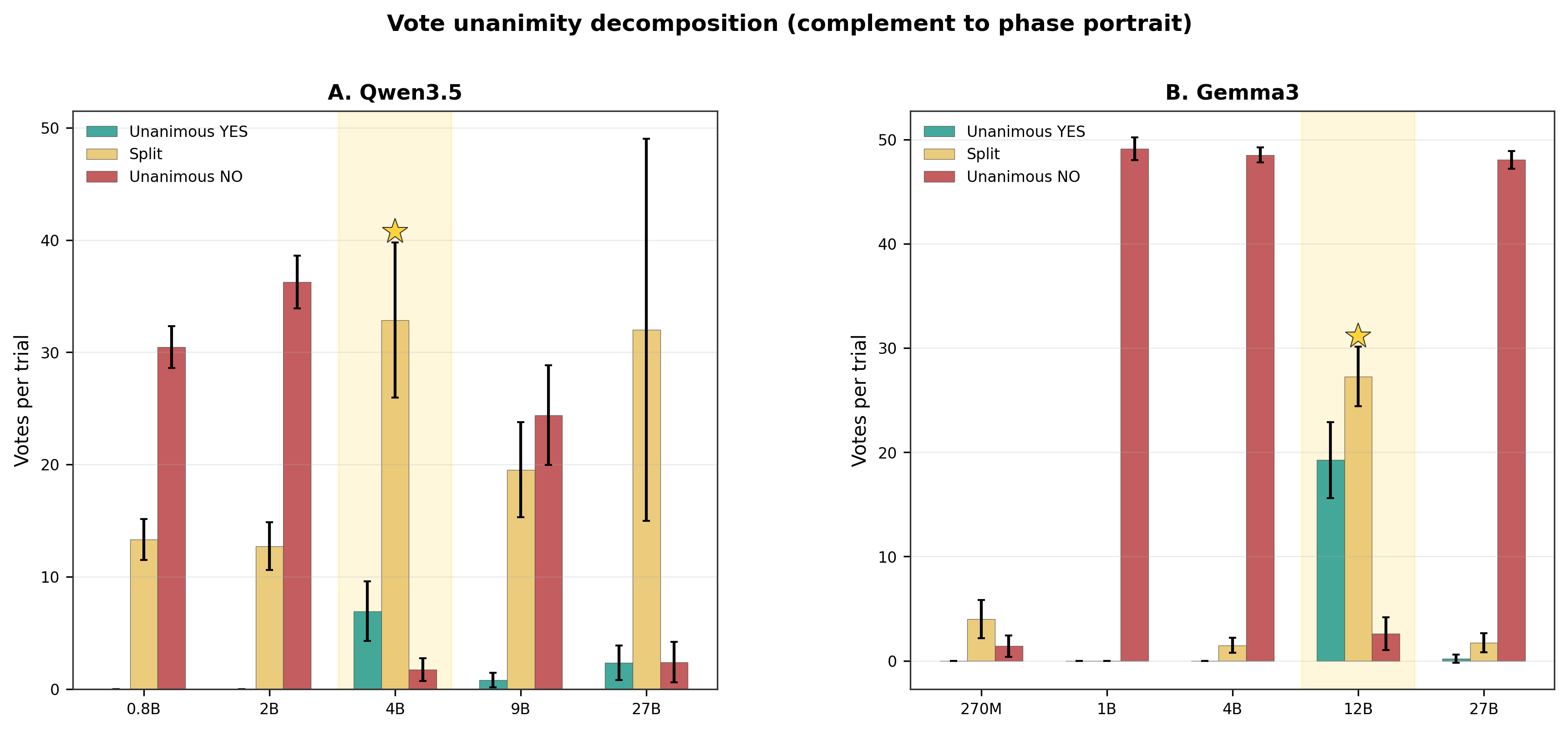}
    \caption{Vote-unanimity decomposition per model scale (left: Qwen3.5; right: Gemma3). Unanimous-approval votes meaningfully exceed unanimous-rejection votes only at the family-specific sweet spot (Qwen3.5 4B; Gemma3 12B); other scales remain rejection-dominated.}
    \label{fig:vote_unanimity_appx}
\end{figure*}

The vote-unanimity decomposition in Figure~\ref{fig:vote_unanimity_appx} separates each adopted vote into unanimous-approval, split, and unanimous-rejection bins.
Only at the family-specific sweet spot (Qwen3.5 4B; Gemma3 12B) does the count of unanimous approvals meaningfully exceed that of unanimous rejections, while smaller and larger scales remain rejection-dominated in both families.
Together with the heatmap above, this confirms that the qualitative shift in collective-decision regime, namely broad targeting combined with balanced approval, occurs only at the sweet spot and is the source of the divergent behavioural endpoints reported in the main text.

\section{Rule-Trajectory Figures}\label{app:trajectory-figs}

Section~\ref{sec:trajectories} argues that the regimes differ in the temporal path of institutional change rather than only in summary statistics; the two figures here visualize that argument directly.
Both panels use the canonical model-colour palette of the main text so that traces can be matched across figures.

\begin{figure*}[h]
    \centering
    \includegraphics[width=\textwidth]{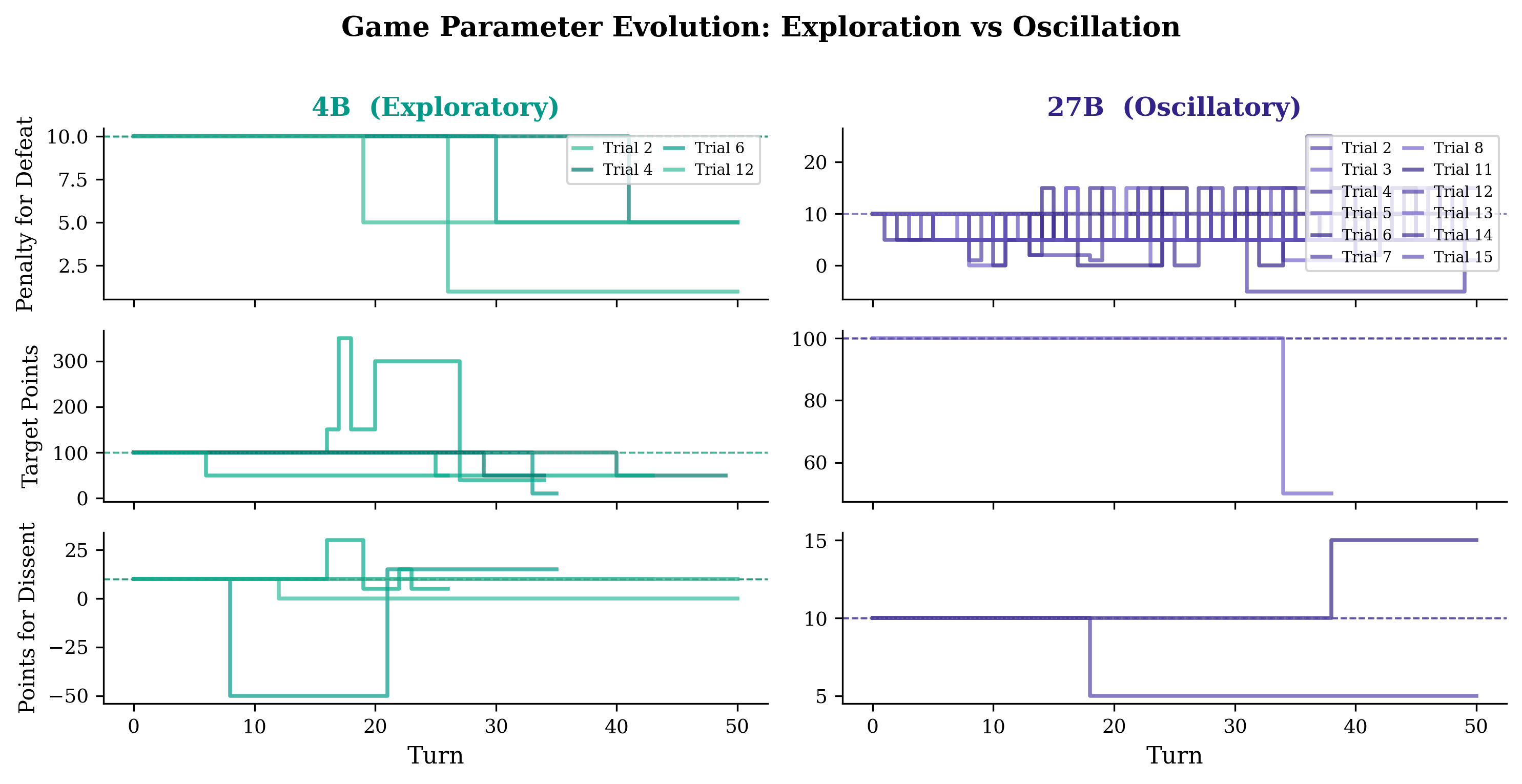}
    \caption{Per-turn trajectories of two patchable fields of Rule~206 at the Qwen3.5 sweet spot (4B) versus the largest scale (27B). 4B exhibits directed exploration followed by stabilisation, whereas 27B cycles through repeatedly visited values ($\sim$79\% reversal rate).}
    \label{fig:rule_trajectories_parameters_appx}
\end{figure*}

Figure~\ref{fig:rule_trajectories_parameters_appx} contrasts the patchable parameters of Rule~206 at the Qwen3.5 sweet spot with those at the largest scale.
The 4B trajectory shows a small number of large adopted moves during the first $\sim$15 turns and then settles near a working point, while the 27B trajectory cycles repeatedly through previously visited values, producing the high reversal rate ($\sim$79\%) reported in Section~\ref{sec:trajectories}.
The two traces therefore correspond to qualitatively different collective control problems, namely convergence at the sweet spot versus chronic non-convergence at the largest scale, rather than to merely different summary statistics.

\begin{figure*}[h]
    \centering
    \includegraphics[width=\textwidth]{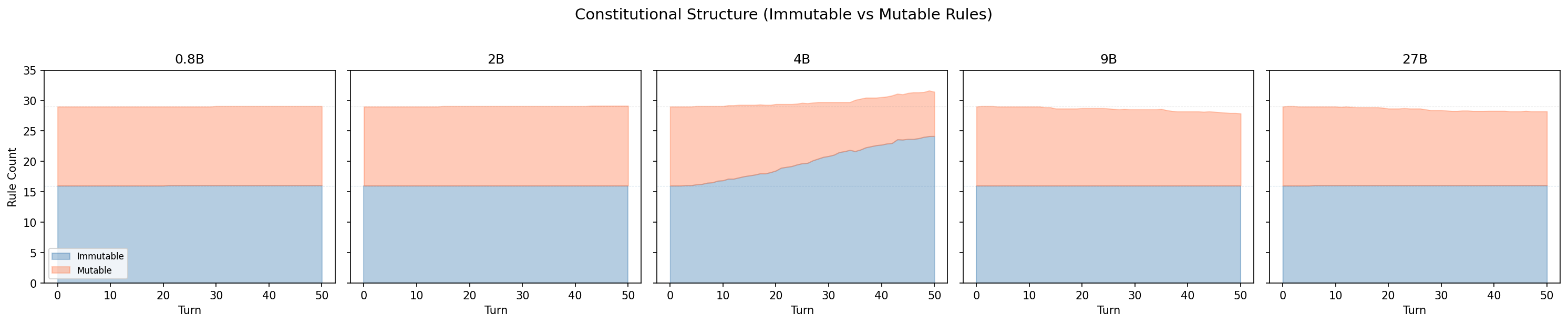}
    \caption{Trajectory of the active-rule count, split into mutable and immutable parts, over the 50-turn game per scale. The sweet spot exhibits a controlled expansion of the mutable subset; frozen scales remain at the 13/16 baseline; larger scales also stay close to baseline because adopted changes are concentrated on AMEND operations.}
    \label{fig:rule_trajectories_mutability_appx}
\end{figure*}

Figure~\ref{fig:rule_trajectories_mutability_appx} adds the constitutional view by tracking the active-rule count split into mutable and immutable parts.
At the frozen scales, both counts remain at the initial 13/16 split.
At the sweet spot the mutable count rises gradually as new mutable rules are added or as immutable rules are transmuted to mutable status, while the immutable count stays near its starting value, a controlled expansion of the legislative space rather than a rewrite of constitutional foundations.
At the largest scales, the count remains close to baseline because adopted changes concentrate on AMEND operations on existing fields rather than on ADD, REPEAL, or TRANSMUTE.
Read together with the parameter trajectories above, this view shows that the regime taxonomy applies not only to which rules change but to how the constitutional structure itself evolves.

\section{\texttt{majority\_from\_start} Intervention: Per-Scale Numerical Detail}\label{app:robustness}

This appendix lists the per-scale numerical breakdown of the \texttt{majority\_from\_start} intervention summarised in Section~\ref{sec:robustness}.
The intervention covers all model scales in both families, 15 trials each, using the same Nomic engine.

Under \texttt{majority\_from\_start}, the Qwen3.5 4B endpoint active-rule count shifts from $29.6$ to $30.4$ ($+2.7\%$) and the mutable-rule fraction from $0.306$ to $0.316$ ($+2.9\%$), both inside baseline between-trial variability, and the ADD/AMEND/REPEAL/TRANSMUTE mix is preserved.
Per-round proposal acceptance rises from $0.356$ to $0.437$ during the unanimity-free phase, but the winner rate drops from 8/15 to 1/15: with unanimity removed from turn~0, premature \texttt{target\_points} amendments are adopted before the player population stabilises, which destabilises convergence at the sweet spot.
At Qwen3.5 27B, the looser threshold reveals five winners out of 15 (vs.\ 1/15 at baseline) entirely through aggressive amendment of Rule~208 to lower \texttt{target\_points} (final values 20--53), consistent with the model's narrow-AMEND repertoire rather than with a transition to varied rule dynamics.
In Gemma3, the 12B model maintains roughly 50\% adoption (49.9\%) and its winner rate rises from 3/15 to 6/15 with \texttt{target\_points} unchanged in every winning trial, so the sweet spot becomes more productive once unanimity is removed.
The Qwen3.5 0.8B and Gemma3 1B models are inert under both conditions, with active rule counts pinned at 29 and adoption rates $\approx 0\%$, and Qwen3.5 2B closely tracks this inert pattern.
The Gemma3 270M model adopts only 1.2\% of all turns at baseline and 1.9\% under intervention (with 89--90\% of turns rejected by the validator before reaching a vote in both conditions), so its constitutional structure remains static regardless of voting threshold.
Consistent with Section~\ref{sec:robustness}, no scale in either family transitions out of its baseline regime under the intervention; the additional Qwen wins are mechanically valid but driven by the same narrow amendment behaviour as at baseline.

\section{Metric Definitions}\label{app:metrics}

This appendix defines the derived metrics used in the main text.

\paragraph{Reversal rate.}
For each model scale, we collect every adopted ENGINE\_PATCH across all trials and group the patched values by parameter name.
A \emph{reversal} occurs when a parameter is set to a value that it has previously held at any earlier point in the game (across all trials for that model scale).
Formally, for each parameter~$p$ we maintain a set~$S_p$ of values observed so far.
When a new adopted patch sets $p \leftarrow v$: if $v \in S_p$, the change is counted as a reversal; then $v$ is added to $S_p$ regardless.
The reversal rate is $R / N \times 100\%$, where $R$~is the total number of reversals and $N$~is the total number of adopted ENGINE\_PATCH changes.

\paragraph{Shannon entropy of rule targeting ($H$).}
Let $f_i$ be the fraction of valid proposals that target rule~$i$.
The entropy is $H = -\sum_i f_i \log_2 f_i$, computed over all rules targeted at least once.
Higher entropy indicates more diverse rule targeting.

\paragraph{EP\% (Engine Patch percentage).}
The fraction of valid proposals that include at least one ENGINE\_PATCH attribute, regardless of whether the proposal is adopted.

\section{Prompt Templates}\label{app:prompts}

This appendix reproduces the exact prompts used for rule-change proposal generation and voting.

\subsection{Proposal Prompt}\label{app:proposal-prompt}

The system preamble is:

\begin{quote}
\small
\texttt{You are Player \{agent\_index\} in a game of Nomic.}

\texttt{Nomic is a game where players take turns proposing rule changes, and all players vote on each proposal.
You are a strategic agent trying to win the game by accumulating points.}

\texttt{Key rules:}\\
\texttt{- You may propose one rule change per turn: ADD, AMEND, REPEAL, or TRANSMUTE.}\\
\texttt{- After proposing and voting, you roll one die and add the result to your score.}\\
\texttt{- If your proposal fails, you lose 10 points (Rule 206).}\\
\texttt{- Players who vote AGAINST a successful proposal gain 10 bonus points (Rule 204, only after unanimity is relaxed).}\\
\texttt{- The first player to reach the target points (default 100) wins.}

\texttt{Think strategically about which rule change will benefit you the most.}
\end{quote}

\noindent\textbf{Note:} The prompt makes the dice-roll mechanism (Rule~202) and the conditional dissent bonus (Rule~204) explicit, so the asymmetric incentive between proposers (no direct bonus for a passing proposal) and dissenters (a $+10$ bonus once unanimity is relaxed) is part of the agents' instruction set.

The user prompt provides the current game state and enforces a structured output format (including rule text, scores, and turn number) with templates for ADD, AMEND, REPEAL, and TRANSMUTE operations.
A patchable attribute map is appended to inform agents which rule parameters can be mechanically modified through an \texttt{ENGINE\_PATCH} block.

\subsection{Voting Prompt}\label{app:vote-prompt}

The voting prompt provides the proposal details and enforces a binary vote (AGREE/DISAGREE) with a reason.


\section{Cross-Family Supplementary Statistics}\label{app:cross_family_stats}

This appendix provides per-condition statistics for both families.

\begin{table}[h]
\centering
\small
\caption{Qwen3.5 baseline behavioural statistics (15 trials per size, temperature~$=0.7$).}
\label{tab:qwen35_baseline}
\begin{tabular}{lccccc}
\toprule
\textbf{Size} & \textbf{Adoption} & \textbf{Winners} & \textbf{Vote YES} & \textbf{Mean Score} & \textbf{Regime} \\
\midrule
0.8B & 0.3\% & 0/15 & 6.8\% & --- & Frozen \\
2B   & 0.3\% & 0/15 & 6.0\% & --- & Frozen \\
4B   & 33.8\% & 8/15 & 59.6\% & --- & Exploratory (sweet spot) \\
9B   & 8.0\% & 0/15 & 17.1\% & --- & Rejection-dominated \\
27B  & 25.6\% & 1/15 & 46.8\% & --- & Narrow-AMEND \\
\bottomrule
\end{tabular}
\end{table}

\begin{table}[h]
\centering
\small
\caption{Gemma3 baseline behavioural statistics (15 trials per size, temperature~$=0.7$).}
\label{tab:gemma3_baseline}
\begin{tabular}{lccccc}
\toprule
\textbf{Size} & \textbf{Adoption} & \textbf{Winners} & \textbf{Vote YES} & \textbf{Mean Score} & \textbf{Regime} \\
\midrule
270M & 1.2\% & 0/15 & --- & --- & Frozen (w/ amendments) \\
1B   & 0.0\%  & 0/15 & $<$2\% & $-$64 & Frozen \\
4B   & 0.1\%  & 0/15 & $<$2\% & $-$64 & Frozen \\
12B  & 53.9\% & 3/15 & 61.6\% & $+$17.5 & Exploratory (sweet spot) \\
27B  & 1.9\%  & 0/15 & $\sim$5\% & $-$61 & Veto-heavy \\
\bottomrule
\end{tabular}
\end{table}

\begin{table}[h]
\centering
\small
\caption{Cross-family \texttt{majority\_from\_start} intervention (15 trials per size).}
\label{tab:cross_family_intervention}
\begin{tabular}{llccc}
\toprule
\textbf{Family} & \textbf{Size} & \textbf{Adoption} & \textbf{Winners} & \textbf{Regime change?} \\
\midrule
Qwen3.5 & 0.8B & $\sim$0\% & 0/15 & No \\
Qwen3.5 & 2B   & $\sim$0\% & 0/15 & No \\
Qwen3.5 & 4B   & $\sim$40\% & 1/15 & No \\
Qwen3.5 & 9B   & $\sim$10\% & 0/15 & No \\
Qwen3.5 & 27B  & $\sim$42\% & 5/15 & Yes (target lowered) \\
\midrule
Gemma3  & 270M & $\sim$2\% & 0/15 & No \\
Gemma3  & 1B   & $\sim$0\% & 0/15 & No \\
Gemma3  & 4B   & $\sim$0\% & 0/15 & No \\
Gemma3  & 12B  & $\sim$50\% & 6/15 & No (enhanced) \\
Gemma3  & 27B  & $\sim$2\% & 0/15 & No \\
\bottomrule
\end{tabular}
\end{table}

\begin{table}[h]
\centering
\small
\caption{Gemma3 temperature sweep summary for 12B (sweet spot). Six trials per temperature; other sizes remain frozen across all temperatures. Mean Score is reported as team total (sum over five players), consistent with the within-table scaling of mean-score values.}
\label{tab:gemma3_temperature}
\begin{tabular}{lccc}
\toprule
\textbf{Temperature} & \textbf{Winners} & \textbf{Adoption} & \textbf{Mean team score} \\
\midrule
0.1 & 0/6 & 39.7\% & $-$72 \\
0.3 & 0/6 & 44.0\% & $-$15 \\
0.5 & 0/6 & 49.3\% & $+$3 \\
0.7 & 1/6 & 53.0\% & $+$36 \\
0.9 & 2/6 & 56.7\% & $+$116 \\
1.1 & 4/6 & 60.5\% & $+$231 \\
1.3 & 4/6 & 62.7\% & $+$237 \\
\bottomrule
\end{tabular}
\end{table}

\paragraph{Heterogeneous societies.}
Both families produce gridlock in mixed-size societies. The Qwen3.5 heterogeneous condition (mixed 0.8B/2B/4B/9B/27B) yields 5.9\% adoption with 0/15 winners and 88.5\% NO votes from the 9B member; the Gemma3 heterogeneous condition (mixed 270M/1B/4B/12B/27B) yields 0.0\% adoption with 0/15 winners and 99.5\% NO votes from the 27B member. The cross-family pattern is therefore the same: the largest member's veto suppresses rule adoption regardless of the presence of a sweet-spot member, with the effect being more severe in Gemma3.

\end{appendices}


\clearpage
\bibliography{bibliography}

@article{hota2025nomiclaw,
  title={NomicLaw: Emergent Trust and Strategic Argumentation in LLMs During Collaborative Law-Making},
  author={Hota, Asutosh and Jokinen, Jussi PP},
  journal={arXiv preprint arXiv:2508.05344},
  year={2025}
}

@book{suber1990paradox,
  title     = {The Paradox of Self-Amendment: A Study of Logic, Law, Omnipotence, and Change},
  author    = {Suber, Peter},
  year      = {1990},
  publisher = {Peter Lang Publishing}
}

@article{kaplan2020scaling,
  title={Scaling laws for neural language models},
  author={Kaplan, Jared and McCandlish, Sam and Henighan, Tom and Brown, Tom B and Chess, Benjamin and Child, Rewon and Gray, Scott and Radford, Alec and Wu, Jeffrey and Amodei, Dario},
  journal={arXiv preprint arXiv:2001.08361},
  year={2020}
}

@article{hoffmann2022training,
  title={Training compute-optimal large language models},
  author={Hoffmann, Jordan and Borgeaud, Sebastian and Mensch, Arthur and Buchatskaya, Elena and Cai, Trevor and Rutherford, Eliza and Casas, DDL and Hendricks, Lisa Anne and Welbl, Johannes and Clark, Aidan and others},
  journal={arXiv preprint arXiv:2203.15556},
  volume={10},
  year={2022}
}

@article{wei2022emergent,
  title={Emergent abilities of large language models},
  author={Wei, Jason and Tay, Yi and Bommasani, Rishi and Raffel, Colin and Zoph, Barret and Borgeaud, Sebastian and Yogatama, Dani and Bosma, Maarten and Zhou, Denny and Metzler, Donald and others},
  journal={Transactions on Machine Learning Research},
  year={2022}
}

@article{meta2022human,
  title={Human-level play in the game of diplomacy by combining language models with strategic reasoning},
  author={Meta Fundamental AI Research Diplomacy Team (FAIR)† and Bakhtin, Anton and Brown, Noam and Dinan, Emily and Farina, Gabriele and Flaherty, Colin and Fried, Daniel and Goff, Andrew and Gray, Jonathan and Hu, Hengyuan and others},
  journal={Science},
  volume={378},
  number={6624},
  pages={1067--1074},
  year={2022},
  publisher={American Association for the Advancement of Science}
}

@article{xu2023exploring,
  title={Exploring large language models for communication games: An empirical study on werewolf},
  author={Xu, Yuzhuang and Wang, Shuo and Li, Peng and Luo, Fuwen and Wang, Xiaolong and Liu, Weidong and Liu, Yang},
  journal={arXiv preprint arXiv:2309.04658},
  year={2023}
}

@article{light2023avalonbench,
  title={Avalonbench: Evaluating llms playing the game of avalon},
  author={Light, Jonathan and Cai, Min and Shen, Sheng and Hu, Ziniu},
  journal={arXiv preprint arXiv:2310.05036},
  year={2023}
}

@article{brown2020language,
  title={Language models are few-shot learners},
  author={Brown, Tom and Mann, Benjamin and Ryder, Nick and Subbiah, Melanie and Kaplan, Jared D and Dhariwal, Prafulla and Neelakantan, Arvind and Shyam, Pranav and Sastry, Girish and Askell, Amanda and others},
  journal={Advances in neural information processing systems},
  volume={33},
  pages={1877--1901},
  year={2020}
}

@misc{qwen2026qwen35,
  title  = {Qwen3.5: Towards Native Multimodal Agents},
  author = {{Qwen Team}},
  month  = {February},
  year   = {2026},
  url    = {https://qwen.ai/blog?id=qwen3.5}
}

@article{flavell1979metacognition,
  title={Metacognition and cognitive monitoring: A new area of cognitive--developmental inquiry.},
  author={Flavell, John H},
  journal={American psychologist},
  volume={34},
  number={10},
  pages={906},
  year={1979},
  publisher={American Psychological Association}
}

@article{didolkar2024metacognitive,
  title={Metacognitive capabilities of llms: An exploration in mathematical problem solving},
  author={Didolkar, Aniket and Goyal, Anirudh and Ke, Nan R and Guo, Siyuan and Valko, Michal and Lillicrap, Timothy and Rezende, Danilo and Bengio, Yoshua and Mozer, Michael and Arora, Sanjeev},
  journal={Advances in Neural Information Processing Systems},
  volume={37},
  pages={19783--19812},
  year={2024}
}

@article{schaeffer2023emergent,
  title={Are emergent abilities of large language models a mirage?},
  author={Schaeffer, Rylan and Miranda, Brando and Koyejo, Sanmi},
  journal={Advances in neural information processing systems},
  volume={36},
  pages={55565--55581},
  year={2023}
}

@inproceedings{park2023generative,
  title={Generative agents: Interactive simulacra of human behavior},
  author={Park, Joon Sung and O'Brien, Joseph and Cai, Carrie Jun and Morris, Meredith Ringel and Liang, Percy and Bernstein, Michael S},
  booktitle={Proceedings of the 36th annual acm symposium on user interface software and technology},
  pages={1--22},
  year={2023}
}

@article{akata2025playing,
  title={Playing repeated games with large language models},
  author={Akata, Elif and Schulz, Lion and Coda-Forno, Julian and Oh, Seong Joon and Bethge, Matthias and Schulz, Eric},
  journal={Nature Human Behaviour},
  volume={9},
  number={7},
  pages={1380--1390},
  year={2025},
  publisher={Nature Publishing Group UK London}
}

@article{gandhi2023strategic,
  title={Strategic reasoning with language models},
  author={Gandhi, Kanishk and Sadigh, Dorsa and Goodman, Noah D},
  journal={arXiv preprint arXiv:2305.19165},
  year={2023}
}

@article{yao2023tree,
  title={Tree of thoughts: Deliberate problem solving with large language models},
  author={Yao, Shunyu and Yu, Dian and Zhao, Jeffrey and Shafran, Izhak and Griffiths, Tom and Cao, Yuan and Narasimhan, Karthik},
  journal={Advances in neural information processing systems},
  volume={36},
  pages={11809--11822},
  year={2023}
}

@inproceedings{horibe2025evolvability,
  title={Evolvability in rule-making: A self-amendment game among llm agents},
  author={Horibe, Kazuya},
  booktitle={Proceedings of the Genetic and Evolutionary Computation Conference Companion},
  pages={2127--2137},
  year={2025}
}

@inproceedings{pan2023rewards,
  title={Do the rewards justify the means? measuring trade-offs between rewards and ethical behavior in the machiavelli benchmark},
  author={Pan, Alexander and Chan, Jun Shern and Zou, Andy and Li, Nathaniel and Basart, Steven and Woodside, Thomas and Zhang, Hanlin and Emmons, Scott and Hendrycks, Dan},
  booktitle={International conference on machine learning},
  pages={26837--26867},
  year={2023},
  organization={PMLR}
}

@inproceedings{lan2024llm,
  title={Llm-based agent society investigation: Collaboration and confrontation in avalon gameplay},
  author={Lan, Yihuai and Hu, Zhiqiang and Wang, Lei and Wang, Yang and Ye, Deheng and Zhao, Peilin and Lim, Ee-Peng and Xiong, Hui and Wang, Hao},
  booktitle={Proceedings of the 2024 Conference on Empirical Methods in Natural Language Processing},
  pages={128--145},
  year={2024}
}

@article{guo2023suspicion,
  title={Suspicion-agent: Playing imperfect information games with theory of mind aware gpt-4},
  author={Guo, Jiaxian and Yang, Bo and Yoo, Paul and Lin, Bill Yuchen and Iwasawa, Yusuke and Matsuo, Yutaka},
  journal={arXiv preprint arXiv:2309.17277},
  year={2023}
}

@article{hu2024survey,
  title={A survey on large language model-based game agents},
  author={Hu, Sihao and Huang, Tiansheng and Liu, Gaowen and Kompella, Ramana Rao and Ilhan, Fatih and Tekin, Selim Furkan and Xu, Yichang and Yahn, Zachary and Liu, Ling},
  journal={arXiv preprint arXiv:2404.02039},
  year={2024}
}

@article{wei2022chain,
  title={Chain-of-thought prompting elicits reasoning in large language models},
  author={Wei, Jason and Wang, Xuezhi and Schuurmans, Dale and Bosma, Maarten and Xia, Fei and Chi, Ed and Le, Quoc V and Zhou, Denny and others},
  journal={Advances in neural information processing systems},
  volume={35},
  pages={24824--24837},
  year={2022}
}

@article{kojima2022large,
  title={Large language models are zero-shot reasoners},
  author={Kojima, Takeshi and Gu, Shixiang Shane and Reid, Machel and Matsuo, Yutaka and Iwasawa, Yusuke},
  journal={Advances in neural information processing systems},
  volume={35},
  pages={22199--22213},
  year={2022}
}

@article{shinn2023reflexion,
  title={Reflexion: Language agents with verbal reinforcement learning},
  author={Shinn, Noah and Cassano, Federico and Gopinath, Ashwin and Narasimhan, Karthik and Yao, Shunyu},
  journal={Advances in neural information processing systems},
  volume={36},
  pages={8634--8652},
  year={2023}
}

@article{madaan2023self,
  title={Self-refine: Iterative refinement with self-feedback},
  author={Madaan, Aman and Tandon, Niket and Gupta, Prakhar and Hallinan, Skyler and Gao, Luyu and Wiegreffe, Sarah and Alon, Uri and Dziri, Nouha and Prabhumoye, Shrimai and Yang, Yiming and others},
  journal={Advances in neural information processing systems},
  volume={36},
  pages={46534--46594},
  year={2023}
}

@article{wang2022self,
  title={Self-consistency improves chain of thought reasoning in language models},
  author={Wang, Xuezhi and Wei, Jason and Schuurmans, Dale and Le, Quoc and Chi, Ed and Narang, Sharan and Chowdhery, Aakanksha and Zhou, Denny},
  journal={arXiv preprint arXiv:2203.11171},
  year={2022}
}

@article{hughes2024open,
  title={Open-endedness is essential for artificial superhuman intelligence},
  author={Hughes, Edward and Dennis, Michael and Parker-Holder, Jack and Behbahani, Feryal and Mavalankar, Aditi and Shi, Yuge and Schaul, Tom and Rocktaschel, Tim},
  journal={arXiv preprint arXiv:2406.04268},
  year={2024}
}

@incollection{lehman2023evolution,
  title={Evolution through large models},
  author={Lehman, Joel and Gordon, Jonathan and Jain, Shawn and Ndousse, Kamal and Yeh, Cathy and Stanley, Kenneth O},
  booktitle={Handbook of evolutionary machine learning},
  pages={331--366},
  year={2023},
  publisher={Springer}
}

@book{stanley2015greatness,
  title={Why Greatness Cannot Be Planned: The Myth of the Objective},
  author={Stanley, Kenneth O. and Lehman, Joel},
  year={2015},
  publisher={Springer}
}

@article{clune2019ai,
  title={AI-GAs: AI-generating algorithms, an alternate paradigm for producing general artificial intelligence},
  author={Clune, Jeff},
  journal={arXiv preprint arXiv:1905.10985},
  year={2019}
}

@article{fernando2023promptbreeder,
  title={Promptbreeder: Self-referential self-improvement via prompt evolution},
  author={Fernando, Chrisantha and Banarse, Dylan and Michalewski, Henryk and Osindero, Simon and Rockt{\"a}schel, Tim},
  journal={arXiv preprint arXiv:2309.16797},
  year={2023}
}

@article{wang2023voyager,
  title={Voyager: An open-ended embodied agent with large language models},
  author={Wang, Guanzhi and Xie, Yuqi and Jiang, Yunfan and Mandlekar, Ajay and Xiao, Chaowei and Zhu, Yuke and Fan, Linxi and Anandkumar, Anima},
  journal={arXiv preprint arXiv:2305.16291},
  year={2023}
}

@article{li2023camel,
  title={Camel: Communicative agents for" mind" exploration of large language model society},
  author={Li, Guohao and Hammoud, Hasan and Itani, Hani and Khizbullin, Dmitrii and Ghanem, Bernard},
  journal={Advances in neural information processing systems},
  volume={36},
  pages={51991--52008},
  year={2023}
}

@inproceedings{zhao2024competeai,
author = {Zhao, Qinlin and Wang, Jindong and Zhang, Yixuan and Jin, Yiqiao and Zhu, Kaijie and Chen, Hao and Xie, Xing},
title = {CompeteAI: understanding the competition dynamics of large language model-based agents},
year = {2024},
publisher = {JMLR.org},
booktitle = {Proceedings of the 41st International Conference on Machine Learning},
articleno = {2526},
numpages = {16},
location = {Vienna, Austria},
series = {ICML'24}
}

@article{mu2023can,
  title={Can LLMs Follow Simple Rules?},
  author={Mu, Norman and Chen, Sarah and Wang, Zifan and Chen, Sizhe and Karamardian, David and Aljeraisy, Lulwa and Alomair, Basel and Hendrycks, Dan and Wagner, David},
  journal={arXiv preprint arXiv:2311.04235},
  year={2023}
}

@article{kosinski2024evaluating,
  title={Evaluating large language models in theory of mind tasks},
  author={Kosinski, Michal},
  journal={Proceedings of the National Academy of Sciences},
  volume={121},
  number={45},
  pages={e2405460121},
  year={2024},
  publisher={National Academy of Sciences}
}

@article{ullman2023large,
  title={Large language models fail on trivial alterations to theory-of-mind tasks},
  author={Ullman, Tomer},
  journal={arXiv preprint arXiv:2302.08399},
  year={2023}
}

@inproceedings{sap2022neural,
  title={Neural theory-of-mind? on the limits of social intelligence in large lms},
  author={Sap, Maarten and Le Bras, Ronan and Fried, Daniel and Choi, Yejin},
  booktitle={Proceedings of the 2022 conference on empirical methods in natural language processing},
  pages={3762--3780},
  year={2022}
}

@article{bai2022constitutional,
  title={Constitutional ai: Harmlessness from ai feedback},
  author={Bai, Yuntao and Kadavath, Saurav and Kundu, Sandipan and Askell, Amanda and Kernion, Jackson and Jones, Andy and Chen, Anna and Goldie, Anna and Mirhoseini, Azalia and McKinnon, Cameron and others},
  journal={arXiv preprint arXiv:2212.08073},
  year={2022}
}

@article{cobben2026gt,
  title={GT-HarmBench: Benchmarking AI Safety Risks Through the Lens of Game Theory},
  author={Cobben, Pepijn and Huang, Xuanqiang Angelo and Pham, Thao Amelia and Dahlgren, Isabel and Zhang, Terry Jingchen and Jin, Zhijing},
  journal={arXiv preprint arXiv:2602.12316},
  year={2026}
}

@article{duan2024gtbench,
  title={Gtbench: Uncovering the strategic reasoning capabilities of llms via game-theoretic evaluations},
  author={Duan, Jinhao and Zhang, Renming and Diffenderfer, James and Kailkhura, Bhavya and Sun, Lichao and Stengel-Eskin, Elias and Bansal, Mohit and Chen, Tianlong and Xu, Kaidi},
  journal={Advances in Neural Information Processing Systems},
  volume={37},
  pages={28219--28253},
  year={2024}
}

@article{masumoto2005new,
  title={A new formalization of a meta-game using the lambda calculus},
  author={Masumoto, Gen and Ikegami, Takashi},
  journal={BioSystems},
  volume={80},
  number={3},
  pages={219--231},
  year={2005},
  publisher={Elsevier}
}

@article{hatakeyama2009minimum,
  title={Minimum Nomic: a tool for studying rule dynamics},
  author={Hatakeyama, Masaomi and Hashimoto, Takashi},
  journal={Artificial Life and Robotics},
  volume={13},
  number={2},
  pages={500--503},
  year={2009},
  publisher={Springer}
}

@article{sun2025game,
  title={Game theory meets large language models: A systematic survey with taxonomy and new frontiers},
  author={Sun, Haoran and Wu, Yusen and Wang, Peng and Chen, Wei and Cheng, Yukun and Deng, Xiaotie and Chu, Xu},
  journal={arXiv preprint arXiv:2502.09053},
  year={2025}
}

@article{lindsey2025emergent,
  author={Lindsey, Jack},
  title={Emergent Introspective Awareness in Large Language Models},
  journal={Transformer Circuits Thread},
  year={2025},
  url={https://transformer-circuits.pub/2025/introspection/index.html}
}

@article{searle1980minds,
  title={Minds, brains, and programs},
  author={Searle, John R},
  journal={Behavioral and brain sciences},
  volume={3},
  number={3},
  pages={417--424},
  year={1980},
  publisher={Cambridge University Press}
}

@article{ouyang2022training,
  title={Training language models to follow instructions with human feedback},
  author={Ouyang, Long and Wu, Jeffrey and Jiang, Xu and Almeida, Diogo and Wainwright, Carroll and Mishkin, Pamela and Zhang, Chong and Agarwal, Sandhini and Slama, Katarina and Ray, Alex and others},
  journal={Advances in neural information processing systems},
  volume={35},
  pages={27730--27744},
  year={2022}
}

@article{christiano2017deep,
  title={Deep reinforcement learning from human preferences},
  author={Christiano, Paul F and Leike, Jan and Brown, Tom and Martic, Miljan and Legg, Shane and Amodei, Dario},
  journal={Advances in neural information processing systems},
  volume={30},
  year={2017}
}

@book{axelrod1984evolution,
  title     = {The Evolution of Cooperation},
  author    = {Axelrod, Robert},
  publisher = {Basic Books},
  year      = {1984}
}

@book{ostrom1990governing,
  title={Governing the commons: The evolution of institutions for collective action},
  author={Ostrom, Elinor},
  year={1990},
  publisher={Cambridge university press}
}

@article{rafailov2023direct,
  title={Direct preference optimization: Your language model is secretly a reward model},
  author={Rafailov, Rafael and Sharma, Archit and Mitchell, Eric and Manning, Christopher D and Ermon, Stefano and Finn, Chelsea},
  journal={Advances in neural information processing systems},
  volume={36},
  pages={53728--53741},
  year={2023}
}

@inproceedings{perez2023discovering,
  title={Discovering language model behaviors with model-written evaluations},
  author={Perez, Ethan and Ringer, Sam and Lukosiute, Kamile and Nguyen, Karina and Chen, Edwin and Heiner, Scott and Pettit, Craig and Olsson, Catherine and Kundu, Sandipan and Kadavath, Saurav and others},
  booktitle={Findings of the association for computational linguistics: ACL 2023},
  pages={13387--13434},
  year={2023}
}

@inproceedings{du2024improving,
  title={Improving factuality and reasoning in language models through multiagent debate},
  author={Du, Yilun and Li, Shuang and Torralba, Antonio and Tenenbaum, Joshua B and Mordatch, Igor},
  booktitle={Forty-first international conference on machine learning},
  year={2024}
}

@article{costarelli2024gamebench,
  title={Gamebench: Evaluating strategic reasoning abilities of llm agents},
  author={Costarelli, Anthony and Allen, Mat and Hauksson, Roman and Sodunke, Grace and Hariharan, Suhas and Cheng, Carlson and Li, Wenjie and Clymer, Joshua and Yadav, Arjun},
  journal={arXiv preprint arXiv:2406.06613},
  year={2024}
}

@article{mckee2023scaffolding,
  title={Scaffolding cooperation in human groups with deep reinforcement learning},
  author={McKee, Kevin R and Tacchetti, Andrea and Bakker, Michiel A and Balaguer, Jan and Campbell-Gillingham, Lucy and Everett, Richard and Botvinick, Matthew},
  journal={Nature Human Behaviour},
  volume={7},
  number={10},
  pages={1787--1796},
  year={2023},
  publisher={Nature Publishing Group UK London}
}

@article{galesic2023beyond,
  title={Beyond collective intelligence: Collective adaptation},
  author={Galesic, Mirta and Barkoczi, Daniel and Berdahl, Andrew M and Biro, Dora and Carbone, Giuseppe and Giannoccaro, Ilaria and Goldstone, Robert L and Gonzalez, Cleotilde and Kandler, Anne and Kao, Albert B and others},
  journal={Journal of the Royal Society Interface},
  volume={20},
  number={200},
  pages={20220736},
  year={2023},
  publisher={The Royal Society},
  doi={10.1098/rsif.2022.0736}
}

@article{alain2017understanding,
  title={Understanding intermediate layers using linear classifier probes},
  author={Alain, Guillaume and Bengio, Yoshua},
  journal={arXiv preprint arXiv:1610.01644},
  year={2016}
}

@article{belinkov2022probing,
  title={Probing classifiers: Promises, shortcomings, and advances},
  author={Belinkov, Yonatan},
  journal={Computational Linguistics},
  volume={48},
  number={1},
  pages={207--219},
  year={2022}
}

@misc{nostalgebraist2020logitlens,
  title={interpreting {GPT}: the logit lens},
  author={nostalgebraist},
  year={2020},
  howpublished={\url{https://www.lesswrong.com/posts/AcKRB8wDpdaN6v6ru/interpreting-gpt-the-logit-lens}}
}

@article{belrose2023eliciting,
  title={Eliciting latent predictions from transformers with the tuned lens},
  author={Belrose, Nora and Ostrovsky, Igor and McKinney, Lev and Furman, Zach and Smith, Logan and Halawi, Danny and Biderman, Stella and Steinhardt, Jacob},
  journal={arXiv preprint arXiv:2303.08112},
  year={2023}
}

@article{meng2022locating,
  title={Locating and editing factual associations in gpt},
  author={Meng, Kevin and Bau, David and Andonian, Alex and Belinkov, Yonatan},
  journal={Advances in neural information processing systems},
  volume={35},
  pages={17359--17372},
  year={2022}
}

@article{vig2020investigating,
  title={Investigating gender bias in language models using causal mediation analysis},
  author={Vig, Jesse and Gehrmann, Sebastian and Belinkov, Yonatan and Qian, Sharon and Nevo, Daniel and Singer, Yaron and Shieber, Stuart},
  journal={Advances in neural information processing systems},
  volume={33},
  pages={12388--12401},
  year={2020}
}

@book{henrich2016secret,
  title={The Secret of Our Success: How Culture Is Driving Human Evolution, Domesticating Our Species, and Making Us Smarter},
  author={Henrich, Joseph},
  year={2016},
  publisher={Princeton University Press}
}

@incollection{boyd2008geneculture,
    author = {Boyd, Robert and Richerson, Peter J.},
    editor = {Engel, Christoph and Singer, Wolf},
    isbn = {9780262195805},
    title = {Gene–Culture Coevolution and the Evolution of Social Institutions},
    booktitle = {Better Than Conscious?: Decision Making, the Human Mind, and Implications For Institutions},
    publisher = {The MIT Press},
    year = {2008},
    month = {05},
    doi = {10.7551/mitpress/9780262195805.003.0014},
    url = {https://doi.org/10.7551/mitpress/9780262195805.003.0014},
    eprint = {https://academic.oup.com/mit-press-scholarship-online/book/0/chapter/175363558/chapter-ag-pdf/44740832/book_17678_section_175363558.ag.pdf},
}

@article{grossmann2023ai,
  title={{AI} and the transformation of social science research},
  author={Grossmann, Igor and Feinberg, Matthew and Parker, Dawn C and Christakis, Nicholas A and Tetlock, Philip E and Cunningham, William A},
  journal={Science},
  volume={380},
  number={6650},
  pages={1108--1109},
  year={2023},
  publisher={American Association for the Advancement of Science}
}

@article{bail2024generative,
  title={Can generative {AI} improve social science?},
  author={Bail, Christopher A},
  journal={Proceedings of the National Academy of Sciences},
  volume={121},
  number={21},
  pages={e2314021121},
  year={2024}
}

@article{argyle2023out,
  title={Out of one, many: Using language models to simulate human samples},
  author={Argyle, Lisa P and Busby, Ethan C and Fulda, Nancy and Gubler, Joshua R and Rytting, Christopher and Wingate, David},
  journal={Political Analysis},
  volume={31},
  number={3},
  pages={337--351},
  year={2023}
}

@inproceedings{wu2025ushaped,
  title={U-shaped and inverted-u scaling behind emergent abilities of large language models},
  author={Wu, Tung-Yu and Lo, Melody},
  booktitle={International Conference on Learning Representations},
  volume={2025},
  pages={99426--99458},
  year={2025}
}

@article{mckenzie2024inverse,
  title={Inverse scaling: When bigger isn't better},
  author={McKenzie, Ian R and Lyzhov, Alexander and Pieler, Michael and Parrish, Alicia and Mueller, Aaron and Prabhu, Ameya and McLean, Euan and Kirtland, Aaron and Ross, Alexis and Liu, Alisa and others},
  journal={Transactions on Machine Learning Research},
  year={2023}
}

@article{google2025gemma3,
  title   = {Gemma 3 Technical Report},
  author  = {{Gemma Team, Google DeepMind}},
  journal = {arXiv preprint arXiv:2503.19786},
  year    = {2025}
}

@article{lovato2025governance,
  title   = {Governance as a complex, networked, democratic, satisfiability problem},
  author  = {Lovato, Juniper and Landry, Nicholas and Hebert-Dufresne, Laurent and others},
  journal = {npj Complexity},
  volume  = {2},
  pages   = {14},
  year    = {2025},
  doi     = {10.1038/s44260-025-00041-3}
}

@article{zeng2026toohuman,
  title   = {Too human to model: the uncanny valley of large language models in simulating human systems},
  author  = {Zeng, Yuanzhao and Brown, Calum and Rounsevell, Mark},
  journal = {npj Complexity},
  volume  = {3},
  pages   = {13},
  year    = {2026},
  doi     = {10.1038/s44260-026-00075-1}
}

@article{piatti2024cooperate,
  title={Cooperate or collapse: Emergence of sustainable cooperation in a society of llm agents},
  author={Piatti, Giorgio and Jin, Zhijing and Kleiman-Weiner, Max and Sch{\"o}lkopf, Bernhard and Sachan, Mrinmaya and Mihalcea, Rada},
  journal={Advances in Neural Information Processing Systems},
  volume={37},
  pages={111715--111759},
  year={2024}
}

@article{conitzer2024social,
  title={Social choice should guide ai alignment in dealing with diverse human feedback},
  author={Conitzer, Vincent and Freedman, Rachel and Heitzig, Jobst and Holliday, Wesley H and Jacobs, Bob M and Lambert, Nathan and Moss{\'e}, Milan and Pacuit, Eric and Russell, Stuart and Schoelkopf, Hailey and others},
  journal={arXiv preprint arXiv:2404.10271},
  year={2024}
}

@book{north1990institutions,
  title     = {Institutions, Institutional Change and Economic Performance},
  author    = {North, Douglass C.},
  year      = {1990},
  publisher = {Cambridge University Press}
}

@article{march1991exploration,
  title   = {Exploration and Exploitation in Organizational Learning},
  author  = {March, James G.},
  journal = {Organization Science},
  volume  = {2},
  number  = {1},
  pages   = {71--87},
  year    = {1991}
}

@inproceedings{huang2024collective,
  title={Collective Constitutional {AI}: Aligning a Language Model with Public Input},
  author={Huang, Saffron and Siddarth, Divya and Lovitt, Liane and Liao, Thomas I and Durmus, Esin and Tamkin, Alex and Ganguli, Deep},
  booktitle={Proceedings of the 2024 ACM Conference on Fairness, Accountability, and Transparency},
  pages={1395--1417},
  year={2024},
  doi={10.1145/3630106.3658979}
}

@article{li2023emergent,
  title={Emergent world representations: Exploring a sequence model trained on a synthetic task},
  author={Li, Kenneth and Hopkins, Aspen K and Bau, David and Vi{\'e}gas, Fernanda and Pfister, Hanspeter and Wattenberg, Martin},
  journal={arXiv preprint arXiv:2210.13382},
  year={2022}
}

@inproceedings{gurnee2024language,
  title={Language models represent space and time},
  author={Gurnee, Wes and Tegmark, Max},
  booktitle={International Conference on Learning Representations},
  volume={2024},
  pages={2483--2503},
  year={2024}
}

\end{document}